\documentclass[a4paper,11pt]{article}
\pdfoutput=1 
\usepackage{subcaption} 
\usepackage{jcappub} 
\usepackage{orcidlink}

\usepackage[T1]{fontenc} 
\usepackage{xspace}
\usepackage{lineno} 
\usepackage{hyperref}
\usepackage{graphicx}  
\usepackage{amsmath}   
\usepackage{textcomp}  

\newcommand{\kvec}{{\bf k}}
\newcommand{\Expmodel}{$\texttt{Exp}$}
\newcommand{\Voimodel}{$\texttt{Voigt}$}

\newcommand{\rpar}{r_{\parallel}}
\newcommand{\rperp}{r_{\perp}}
\newcommand{\blya}{b_{\rm Ly\alpha}}
\newcommand{\bquasar}{b_{\rm Q}}
\newcommand{\betaquasar}{\beta_{\rm Q}}
\newcommand{\biasetalya}{b^{\eta}_{\rm Ly\alpha}}
\newcommand{\biasetaHCD}{b^{\eta}_{\rm HCD}}
\newcommand{\betalya}{\beta_{\rm Ly\alpha}}
\newcommand{\bhcd}{b_{\rm HCD}}
\newcommand{\bfhcd}{b^{F}_{\rm HCD}}
\newcommand{\betahcd}{\beta_{\rm HCD}}
\newcommand{\rhohcd}{\rho_{\rm HCD}}
\newcommand{\Fhcd}{F_{\rm HCD}}
\newcommand{\Fhcdvoigt}{F_{\rm HCD}^{\rm Voigt}}
\newcommand{\Lhcd}{L_{\rm HCD}}
\newcommand{\kpar}{k_{\parallel}}
\newcommand{\lcdm}{$\Lambda$CDM}
\newcommand{\lya}{Ly$\alpha$}
\newcommand{\nhi}{N_{\rm{HI}}}
\newcommand{\lrf}{\lambda_{\rm RF}}
\newcommand{\apar}{\alpha_{\parallel}}
\newcommand{\aperp}{\alpha_{\perp}}
\newcommand{\hmpc} {h^{-1}\mathrm{Mpc}}
\newcommand{\lognhi} {\log N_{\rm{HI}}}
\newcommand{\lognhicm} {\log N_{\rm{HI}}/{1\rm{cm}^{-2}}}
\newcommand{\sigmav} {\sigma_v}
\newcommand{\lyaf}{Lyman-$\alpha$ forest\xspace}

\title{\boldmath Modeling of the High Column Density systems in The Lyman-Alpha Forest}

\author[1,2,3]{{T.~Tan}\orcidlink{0000-0001-8289-1481},}
\author[1]{{J.~Rich},}
\author[4]{{E.~Chaussidon}\orcidlink{0000-0001-8996-4874},}
\author[1]{{J.M.~Le~Goff},}
\author[2]{{C.~Balland},}
\author[1]{{E.~Armengaud}\orcidlink{0000-0001-7600-5148},}
\author[4]{{J.~Aguilar},}
\author[5]{{S.~Ahlen}\orcidlink{0000-0001-6098-7247},}
\author[6,7]{{D.~Bianchi}\orcidlink{0000-0001-9712-0006},}
\author[8]{{D.~Brooks},}
\author[4]{{T.~Claybaugh},}
\author[4]{{A.~Cuceu}\orcidlink{0000-0002-2169-0595},}
\author[9]{{A.~de la Macorra}\orcidlink{0000-0002-1769-1640},}
\author[8]{{P.~Doel},}
\author[10,11]{{J.~E.~Forero-Romero}\orcidlink{0000-0002-2890-3725},}
\author[12,13,14]{{E.~Gaztañaga},}
\author[4,15]{{S.~Gontcho A Gontcho}\orcidlink{0000-0003-3142-233X},}
\author[16]{{G.~Gutierrez},}
\author[17,1]{{H.~K.~Herrera-Alcantar}\orcidlink{0000-0002-9136-9609},}
\author[18,19,20]{{K.~Honscheid}\orcidlink{0000-0002-6550-2023},}
\author[21]{{M.~Ishak}\orcidlink{0000-0002-6024-466X},}
\author[22]{{D.~Kirkby}\orcidlink{0000-0002-8828-5463},}
\author[4]{{T.~Kisner}\orcidlink{0000-0003-3510-7134},}
\author[4]{{A.~Kremin}\orcidlink{0000-0001-6356-7424},}
\author[4]{{M.~Landriau}\orcidlink{0000-0003-1838-8528},}
\author[2]{{L.~Le~Guillou}\orcidlink{0000-0001-7178-8868},}
\author[23,24]{{M.~Manera}\orcidlink{0000-0003-4962-8934},}
\author[18,25,20]{{P.~Martini}\orcidlink{0000-0002-4279-4182},}
\author[26,24]{{R.~Miquel},}
\author[13]{{S.~Nadathur}\orcidlink{0000-0001-9070-3102},}
\author[1,4]{{N.~Palanque-Delabrouille}\orcidlink{0000-0003-3188-784X},}
\author[27]{{F.~Prada}\orcidlink{0000-0001-7145-8674},}
\author[28]{{I.~P\'erez-R\`afols}\orcidlink{0000-0001-6979-0125},}
\author[29]{{G.~Rossi},}
\author[30]{{E.~Sanchez}\orcidlink{0000-0002-9646-8198},}
\author[4]{{D.~Schlegel},}
\author[31,32]{{M.~Schubnell},}
\author[33]{{H.~Seo}\orcidlink{0000-0002-6588-3508},}
\author[4]{{J.~Silber}\orcidlink{0000-0002-3461-0320},}
\author[34]{{D.~Sprayberry},}
\author[32]{{G.~Tarl\'{e}}\orcidlink{0000-0003-1704-0781},}
\author[35,36]{{M.~Walther}\orcidlink{0000-0002-1748-3745},}
\author[34]{{B.~A.~Weaver},}
\author[37]{{H.~Zou}\orcidlink{0000-0002-6684-3997},}

\affiliation[1]{IRFU, CEA, Universit\'{e} Paris-Saclay, F-91191 Gif-sur-Yvette, France}
\affiliation[2]{Sorbonne Universit\'{e}, CNRS/IN2P3, Laboratoire de Physique Nucl\'{e}aire et de Hautes Energies (LPNHE), FR-75005 Paris, France}
\affiliation[3]{CNRS-UCB International Research Laboratory, Centre Pierre Binétruy, IRL2007, CPB-IN2P3, Berkeley, US}
\affiliation[4]{Lawrence Berkeley National Laboratory, 1 Cyclotron Road, Berkeley, CA 94720, USA}
\affiliation[5]{Department of Physics, Boston University, 590 Commonwealth Avenue, Boston, MA 02215 USA}
\affiliation[6]{Dipartimento di Fisica ``Aldo Pontremoli'', Universit\`a degli Studi di Milano, Via Celoria 16, I-20133 Milano, Italy}
\affiliation[7]{INAF-Osservatorio Astronomico di Brera, Via Brera 28, 20122 Milano, Italy}
\affiliation[8]{Department of Physics \& Astronomy, University College London, Gower Street, London, WC1E 6BT, UK}
\affiliation[9]{Instituto de F\'{\i}sica, Universidad Nacional Aut\'{o}noma de M\'{e}xico,  Circuito de la Investigaci\'{o}n Cient\'{\i}fica, Ciudad Universitaria, Cd. de M\'{e}xico  C.~P.~04510,  M\'{e}xico}
\affiliation[10]{Departamento de F\'isica, Universidad de los Andes, Cra. 1 No. 18A-10, Edificio Ip, CP 111711, Bogot\'a, Colombia}
\affiliation[11]{Observatorio Astron\'omico, Universidad de los Andes, Cra. 1 No. 18A-10, Edificio H, CP 111711 Bogot\'a, Colombia}
\affiliation[12]{Institut d'Estudis Espacials de Catalunya (IEEC), c/ Esteve Terradas 1, Edifici RDIT, Campus PMT-UPC, 08860 Castelldefels, Spain}
\affiliation[13]{Institute of Cosmology and Gravitation, University of Portsmouth, Dennis Sciama Building, Portsmouth, PO1 3FX, UK}
\affiliation[14]{Institute of Space Sciences, ICE-CSIC, Campus UAB, Carrer de Can Magrans s/n, 08913 Bellaterra, Barcelona, Spain}
\affiliation[15]{University of Virginia, Department of Astronomy, Charlottesville, VA 22904, USA}
\affiliation[16]{Fermi National Accelerator Laboratory, PO Box 500, Batavia, IL 60510, USA}
\affiliation[17]{Institut d'Astrophysique de Paris. 98 bis boulevard Arago. 75014 Paris, France}
\affiliation[18]{Center for Cosmology and AstroParticle Physics, The Ohio State University, 191 West Woodruff Avenue, Columbus, OH 43210, USA}
\affiliation[19]{Department of Physics, The Ohio State University, 191 West Woodruff Avenue, Columbus, OH 43210, USA}
\affiliation[20]{The Ohio State University, Columbus, 43210 OH, USA}
\affiliation[21]{Department of Physics, The University of Texas at Dallas, 800 W. Campbell Rd., Richardson, TX 75080, USA}
\affiliation[22]{Department of Physics and Astronomy, University of California, Irvine, 92697, USA}
\affiliation[23]{Departament de F\'{i}sica, Serra H\'{u}nter, Universitat Aut\`{o}noma de Barcelona, 08193 Bellaterra (Barcelona), Spain}
\affiliation[24]{Institut de F\'{i}sica d’Altes Energies (IFAE), The Barcelona Institute of Science and Technology, Edifici Cn, Campus UAB, 08193, Bellaterra (Barcelona), Spain}
\affiliation[25]{Department of Astronomy, The Ohio State University, 4055 McPherson Laboratory, 140 W 18th Avenue, Columbus, OH 43210, USA}
\affiliation[26]{Instituci\'{o} Catalana de Recerca i Estudis Avan\c{c}ats, Passeig de Llu\'{\i}s Companys, 23, 08010 Barcelona, Spain}
\affiliation[27]{Instituto de Astrof\'{i}sica de Andaluc\'{i}a (CSIC), Glorieta de la Astronom\'{i}a, s/n, E-18008 Granada, Spain}
\affiliation[28]{Departament de F\'isica, EEBE, Universitat Polit\`ecnica de Catalunya, c/Eduard Maristany 10, 08930 Barcelona, Spain}
\affiliation[29]{Department of Physics and Astronomy, Sejong University, 209 Neungdong-ro, Gwangjin-gu, Seoul 05006, Republic of Korea}
\affiliation[30]{CIEMAT, Avenida Complutense 40, E-28040 Madrid, Spain}
\affiliation[31]{Department of Physics, University of Michigan, 450 Church Street, Ann Arbor, MI 48109, USA}
\affiliation[32]{University of Michigan, 500 S. State Street, Ann Arbor, MI 48109, USA}
\affiliation[33]{Department of Physics \& Astronomy, Ohio University, 139 University Terrace, Athens, OH 45701, USA}
\affiliation[34]{NSF NOIRLab, 950 N. Cherry Ave., Tucson, AZ 85719, USA}
\affiliation[35]{Excellence Cluster ORIGINS, Boltzmannstrasse 2, D-85748 Garching, Germany}
\affiliation[36]{University Observatory, Faculty of Physics, Ludwig-Maximilians-Universit\"{a}t, Scheinerstr. 1, 81677 M\"{u}nchen, Germany}
\affiliation[37]{National Astronomical Observatories, Chinese Academy of Sciences, A20 Datun Road, Chaoyang District, Beijing, 100101, P.~R.~China}

\emailAdd{ting.tan@cea.fr}
\emailAdd{james.rich@cea.fr}

\abstract{
The Lyman-$\alpha$ forests observed in the spectra of high-redshift quasars can be used as a tracer of the cosmological matter density to study baryon acoustic oscillations (BAO) and the Alcock-Paczynski effect. Extraction of cosmological information from these studies  requires modeling of the forest correlations. While the models depend most importantly on the bias parameters of the intergalactic medium (IGM), they also depend on the numbers and characteristics of high-column-density systems (HCDs) ranging from Lyman-limit systems with column densities $\lognhicm>17$
to damped Lyman-$\alpha$ systems (DLAs) with 
$\lognhicm>20.2$.
These HCDs introduce broad damped absorption characteristic of a Voigt profile. Consequently they imprint a component on the power spectrum 
whose modes in the radial direction are suppressed, leading to
a scale-dependent bias.
Using mock data sets of known HCD content,
we test a model that describes this effect in terms of the distribution of column densities of HCDs, the Fourier transforms of their Voigt profiles
and the bias of the halos containing the HCDs.
Our results show that this physically well-motivated model describes the effects of HCDs with an accuracy comparable to that of the ad-hoc  models used in published forest analyses. 
We also discuss the problems of applying the model to real data, where the HCD content and their bias is uncertain.
}

\begin{document}
\maketitle
\flushbottom
\section{Introduction}
Lyman-$\alpha$ forests, detected as a wavelength-dependent absorption 
in quasar spectra, can be used as biased tracers for the underlying dark matter density field. The forests have been used to measure the baryon acoustic oscillations (BAO) at a high redshift $z>2$ 
through the forest auto correlations
\citep{2013BuscaBOSS, 2013SlosarDR9, delubac2015baryon, bautista2017measurement, de2019baryon}, 
quasar-forest cross-correlations
\citep{Font+14, 2017duMasDR12, 2019BlomqvistDR14} and the combination of the two \citep{dmdb2020,DESIyr1lyabao,DESI-yr3lyabao}.  
The Alcock-Paczynski effect has also been studied \citep{CuceuAP}.
These studies have used the quasar spectra collected from the BOSS \citep{BOSS2013} and eBOSS programs \citep{dawson2016sdss, alam2021completed}, and the on-going Dark Energy Spectroscopic Instrument (DESI) program \citep{Snowmass2013.Levi,aghamousa2016desi,abareshi2022overview}
described in detail in 
\citep{
DESI2016b.Instr,
FocalPlane.Silber.2023,
FiberSystem.Poppett.2024,
Corrector.Miller.2023, 
Spectro.Pipeline.Guy.2023,
SurveyOps.Schlafly.2023,
DESI2023b.KP1.EDR,DESI2023a.KP1.SV,DESIyr1lyabao,DESI-yr3lyabao,LyaBAO.EDR.Gordon.2023, 2024arXiv240403001D, DESI2024.II.KP3,2024arXiv240403000D,DESI2024.V.KP5,2024arXiv240403002D,DESI2024.VII.KP7B}.

Most forest absorption is due to neutral Hydrogen
in the intergalactic medium (IGM) and
its relation to the underlying matter fluctuations is understood
through the fluctuating Gunn-Peterson approximation (FGPA) \citep{FGPHreference}.
The amplitude of long-range correlations involving the diffuse IGM component of the forest can be modeled 
with the introduction of bias parameters giving the proportionality between forest and matter fluctuations.
Additional absorption due to high-column density systems (HCDs) when the line of sight passes near a galaxy (the circumgalactic medium) or through a galaxy (the interstellar medium) is not caused by the diffuse IGM.
This complicates the forest correlations because their absorption profiles spread out the correlations radially.
In the power spectrum, this suppresses modes of large $\kpar$, the component of $k$ along the line of sight, leading to scale-dependent bias parameters.

The purpose of this paper is to study the modeling of the mode suppression
due to HCDs with column densities $>10^{17.2}{\rm cm^{-2}}$. 
This value of the column density corresponds to an HCD with unit optical depth for photons with wavelengths less than the Lyman-limit, $911.3$~\AA.
HCDs with column densities $>10^{20.2}{\rm cm^{-2}}$ are called damped \lya~systems  (DLAs).
DLAs have absorption features that are sufficiently wide to make them detectable
in the surveys of  SDSS \citep{prochaska2005sdss,noterdaeme2009evolution} including
BOSS and eBOSS \citep{noterdaeme2012column, parks2018deep,garnett2017detecting,fumagalli2020detecting,  chabanier2022completed}, and DESI \citep{wang2022deep,brodzellerdla}.
We will see that DLAs dominate the mode-suppression unless they are eliminated from the forests by masking the wavelength range with large absorption.
If they are masked, the mode-suppression is dominated by HCDs with column densities just
below that of the DLAs, $>10^{19}{\rm cm^{-2}}$.

\begin{figure}
\centering
\includegraphics[width=0.7\textwidth]{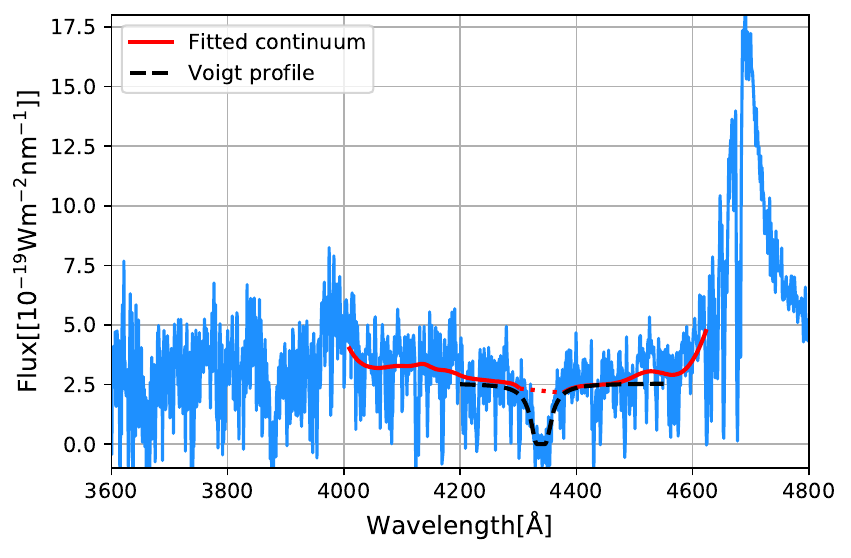}
\caption{A flux spectrum, $f_q(\lambda)$, for a mock quasar, with $z_{\text{QSO}}=2.86$. The \lya~emission line is present  at 4685\AA~and a DLA with $N_{\text{HI}}=10^{20.72}\text{cm}^{-2}$ at redshift $z_{\text{DLA}}=2.57$ is present at 4339.9\AA.
The red line shows the estimated mean spectrum, $\overline{F}(\lambda)C_q(\lambda)$, about which the fluctuations
are measured. The red dotted line indicates the region that is masked in the standard picca analysis. The black line shows the fitted Voigt profile fit to this DLA.
}
\label{fig:forest_spectrum}
\end{figure}

\begin{figure}
\centering
\includegraphics[width=\textwidth]{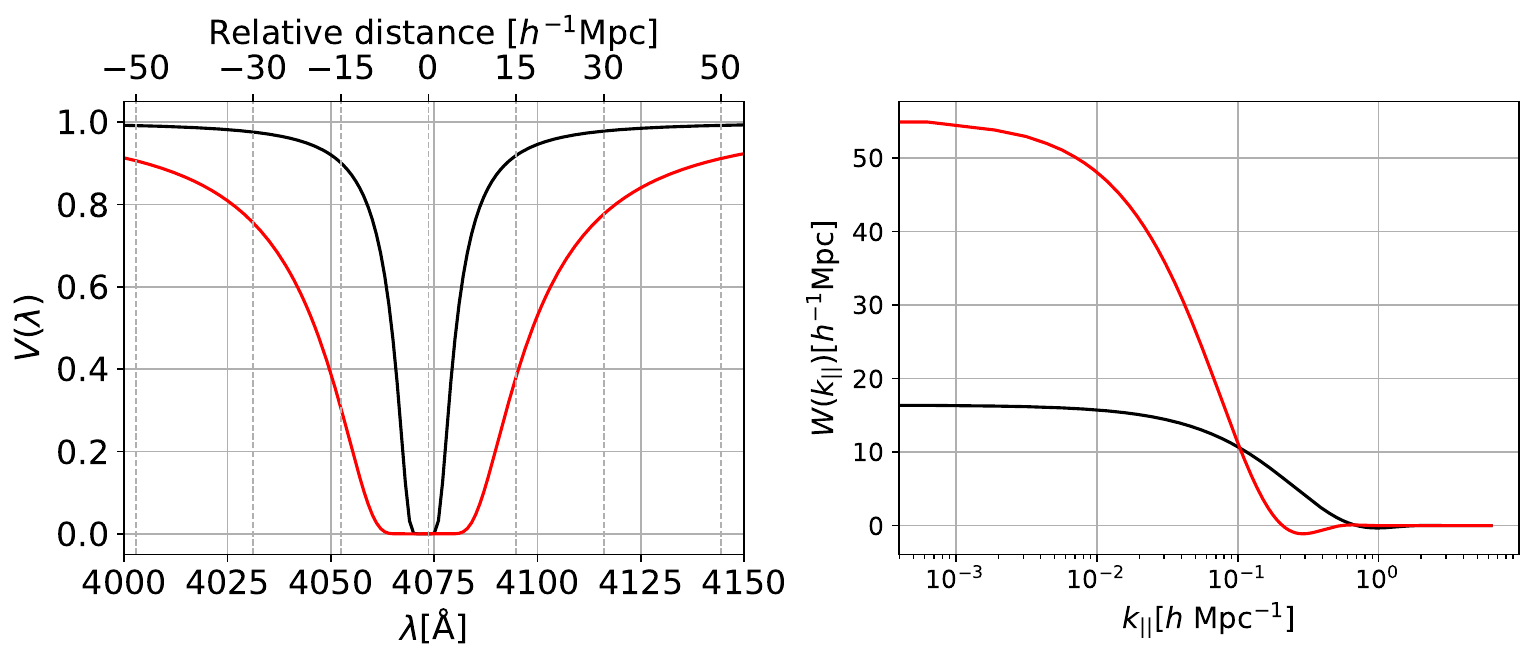}
\caption{The Voigt transmission profile, $V(\lambda)$ and the Fourier transform, $W(k)$ of $W(\lambda)=1-V(\lambda)$ for HCDs with $N_{\text{HI}}=10^{20}\text{cm}^{-2}$ (black) and $10^{21}\text{cm}^{-2}$ (red)
at a redshift of 2.35. The distance and $\kpar$ scales are derived from the wavelength scale using the fiducial cosmological model.
}
\label{fig:voigt_profile}
\end{figure}

An example of a quasar spectrum with a DLA is shown in
Figure~\ref{fig:forest_spectrum}.
The quasar of redshift $z_{\text{QSO}}=2.86$ is taken from mock data sets described in \citet{Etourneau2024}. 
The
fluctuating absorption on the blue side of \lya~emission ($\lambda=$4685\AA) 
is mostly due to HI absorption in the IGM but there is an additional
absorption at $\lambda=$4340\AA~due to 
a DLA of column density
$N_{\text{HI}}=10^{20.72}\text{cm}^{-2}$ at a redshift $z_{\text{HCD}}=2.57$. 
%
As illustrated by this spectrum, DLAs are seen as strong absorption regions 
with damping wings 
in quasar spectra, and can be modeled using the Voigt profile (Fig. \ref{fig:voigt_profile}), a convolution of the Lorentzian absorption profile and Gaussian thermal broadening. 
The bias of DLAs considered as discrete objects is measurable through the 
cross-correlation of DLAs with Ly$\alpha$ forests \citep{font2012hcdforestcorr,perez2018cosmological,2023MNRAS.524.1464P} or through cross-correlation with CMB lensing \citep{alonso2018bias}.
About 6\% for \lya~forests of quasars with redshift $\approx2.5$ contain a DLA in the
wavelength range between the \lya~and Ly$\beta$ emission lines \citep{noterdaeme2012column,chabanier2022completed}.

To minimize the complication of modeling the suppression of large-$\kpar$ modes in the forest power spectrum,
detected DLAs can be masked from quasar spectra, as illustrated in Fig. \ref{fig:forest_spectrum}.
The masking is, however, imperfect, because DLAs in low signal-to-noise forests cannot be detected and masked. Furthermore, 
 HCDs with $N_{\text{HI}}<10^{20.3}{\rm cm^{-2}}$ cannot be detected and  have a non-negligible effect on the Ly$\alpha$ correlation function. A model of this effect on large scales was developed in \citet{McQuinn2011} (noted MW11 hereafter) and \citet{font2012effect} (noted FM12 hereafter) and further developed in \cite{rogers2018correlations}.
These models relate the damping of the radial correlations to the Fourier transform of the Voigt profiles of the HCDs and to the bias of the halos containing the HCDs.
They have not yet been used in the published analyses 
of BOSS, eBOSS and DESI data, which have all modeled the power-spectrum cutoff at high $\kpar$ with functional forms inspired by Ly$\alpha$ forest simulations
\cite{rogers2018correlations,rogers2018simulating}. 

The purpose of this paper is to evaluate the accuracy of the models of MW11 and FM12 in describing the effects of HCDs on forest correlations 
in the case where the numbers and biases of the HCD's are known.
To this end, we use the sets of mock quasar spectra described in \citet{Etourneau2024}. 
These mock spectra were produced with realistic forest auto-correlations and quasar-forest cross-correlations, and with spectral resolution characteristic of the eBOSS and/or DESI programs.  Most importantly, the mocks are produced with HCDs with the expected column-density distribution and with a bias consistent with that measured for DLAs.

This paper is organized as follows. 
In Section \ref{sec::analprocedure}, we describe the procedure to measure the auto-correlations of \lya~absorption and its cross-correlations with quasars.
In Section \ref{sec::model} we describe the theoretical framework of our model of forest correlations. In Section \ref{sec:testswithmocks} we present tests of  the models using mock data. 
Section \ref{sec:eBOSS} presents fits of the eBOSS data and discuss the problems that occur when the numbers of HCD's and their biases are not known.
We conclude  in Section \ref{sec:conclusions}.

\begin{figure}
\begin{minipage}{1.0\textwidth}
    \centering
    \includegraphics[width=\textwidth]{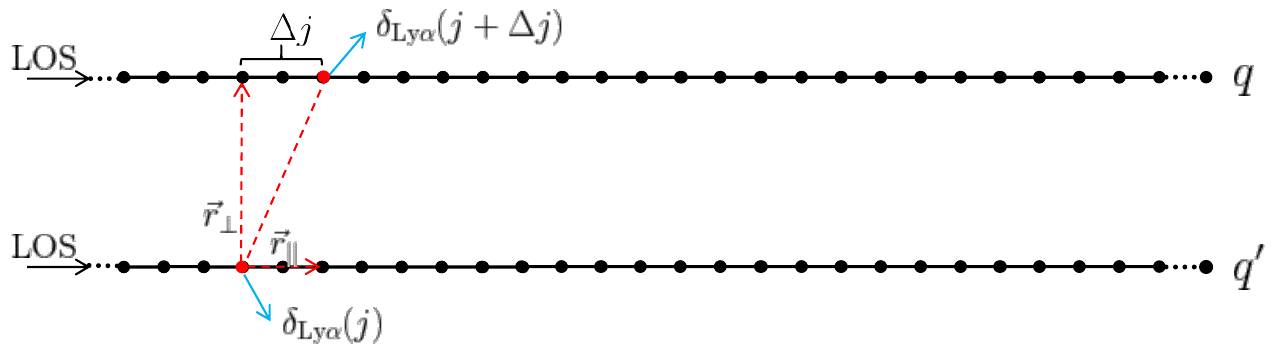}
\end{minipage}

\vspace{20pt} 
\begin{minipage}{1.0\textwidth}
    \centering
    \includegraphics[width=\textwidth]{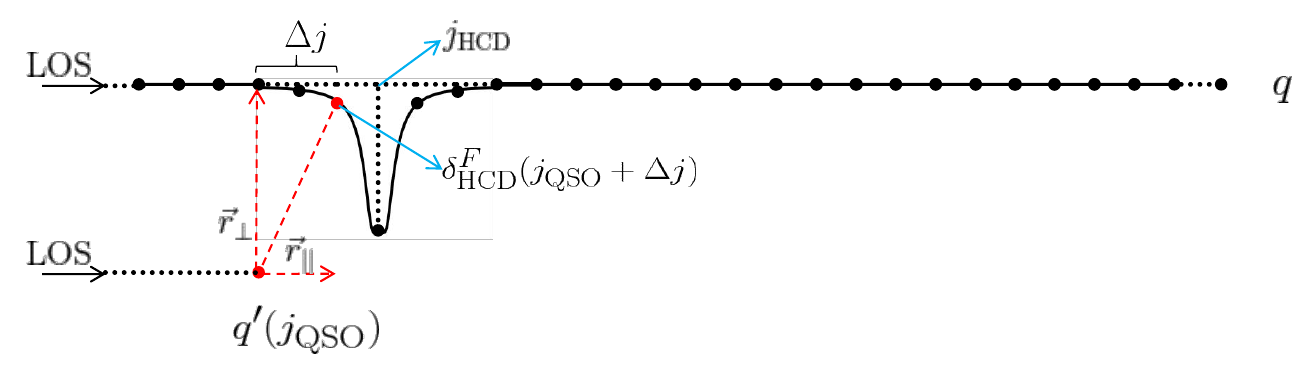}
\end{minipage}
\caption{
Illustration of the flux-flux auto-correlations and quasar-flux cross-correlations.
The top two lines of sight illustrate flux-flux auto-correlation for two forests in front of quasars $q$ and $q'$. The bottom two lines of sight illustrate the quasar-flux cross-correlation.
Flux measurements $f_q(\lambda)$ along the LOS to quasar $q$ and the associated flux-transmission field $\delta_q(\lambda)$ are measured on ``pixels'' of width $\Delta\log\lambda=0.0001$ (BOSS and eBOSS) or $\Delta\lambda=0.8$\AA~(DESI) as represented by the dots and indexed by $j$.  
On the bottom plot a HCD is centered on a wavelength indexed by $j_{HCD}$.
Correlations of $\delta_q(\lambda)$ with a neighboring pixel or quasar are measured as a function of $(\rperp,\rpar)$ calculated from the pixel wavelength separation and angular separation of the two lines-of-sight assuming a fiducial cosmological model.
}
\label{fig:xcf QSO HCD}
\end{figure}

\section{Analysis procedure of forest correlations}
\label{sec::analprocedure}

In this section we list the steps needed to measure the \lya~forest flux transmission field and its auto-correlation and cross-correlation with discrete tracers like quasars.
The implementation of these steps is done with 
the standard eBOSS and DESI pipeline package \texttt{picca}\footnote{https://github.com/igmhub/picca} \cite{du2021picca},
The details of all steps are described for instance in~\cite{dmdb2020}. 

The geometry of the correlation measurements are illustrated in Fig. \ref{fig:xcf QSO HCD}.
The main steps of the measurement are as follows:
 

\begin{enumerate}

\item Establishment of a catalog of quasars, their forests, and DLAs in the forests.
For mocks data, the input list of quasars and their redshifts is known while 
for the data, the quasar catalog is constructed  by algorithms used to determine the nature of the observed spectra (quasar, galaxy, star). 
The forests of each quasar are defined over the restframe wavelength range
$1040<\lrf<1200$\AA.  
We discard observed wavelengths $<3600$\AA~where the CCD throughput is very low.
The DLAs in mock data are known while for the data
DLA-finding algorithms must be used to construct a catalog and determine redshifts and column densities.
The completeness and purity of the resulting DLA catalogs vary with the quasar signal to noise and the system column density.
If a DLA masking strategy is adopted, the range of wavelengths where the DLA transmission less than 0.8 is excluded from the forest.

\item 
Calculation of  
the flux-transmission field, $\delta_q(\lambda)$, 
for each quasar, $q$, at wavelength $\lambda$ from the  measured flux $f_q(\lambda)$
(blue line of Fig \ref{fig:forest_spectrum}):
\begin{equation}
    \delta_{q}(\lambda) =
    \frac{
    f_{q}(\lambda)
    }{
    \overline{F}(\lambda)C_{q}(\lambda)
    } - 1
 \; .
    \label{equation::definition_delta}
\end{equation}
Here $\overline{F}(\lambda)$ is the mean transmitted flux fraction 
and $C_q(\lambda)$ is the estimated
unabsorbed quasar continuum.  
The product $\overline{F}(\lambda)C_q(\lambda)$ is shown as the red line in Fig. \ref{fig:forest_spectrum}.
As illustrated in Fig. \ref{fig:xcf QSO HCD}, the $\delta_q(\lambda)$ are defined on a wavelength grid with $\Delta\log\lambda=0.0001$ for BOSS and eBOSS and $\Delta\lambda=0.8$\AA~for DESI.
In the standard analysis (Sect. 2.4 of \cite{dmdb2020}), 
we use the $\delta_q(\lambda)$ 
in the restframe wavelength range between the \lya~and Ly$\beta$ quasar lines: $1040<\lrf<1200$\AA. 
The product 
$\overline{F}(\lambda)C_{q}(\lambda)$ 
is determined by fitting each forest
to the mean
forest spectrum modified by two free parameters per quasar reflecting the
quasar brightness and spectral index.
The mean forest spectrum  is modified to include the effects of the wings of masked DLAs.

\item Use of a  fiducial cosmological model to transform angular separations and redshift separations
$(\Delta z,\Delta\theta)$ between pairs of flux-transmission field elements $(\delta_q(\lambda),\delta_{q^\prime}(\lambda^\prime))$ into co-moving distances $(\rperp,\rpar)$
(Section 3.1 of \cite{dmdb2020}).
Redshifts are derived from the observed wavelength $\lambda$ assuming 
\lya~absorption.

\item Calculation of the forest auto-correlation function
and quasar-forest cross-correlation 
in bins
$\Delta\rperp=\Delta\rpar=4\hmpc$
(Sections 3.2 and 3.3 of \cite{dmdb2020}):
\begin{equation}
	\xi_{A}^{auto} = \frac{
    \sum\limits_{(i,j) \in A} w_{i}w_{j} \, \delta_{i}\delta_{j}
    }{
	\sum\limits_{(i,j) \in A} w_{i}w_{j}
    }
    \hspace*{10mm}
    \xi_{A}^{cross} = \frac{
    \sum\limits_{(i,q) \in A} w_q w_{i} \, \delta_{i}
    }{
	\sum\limits_{(i,q) \in A} w_q w_{i}
    }.
    \label{equation::xi_estimators]}
\end{equation}
Here, $i$ and $j$ represent pixels on the $\lambda$ grid of a given quasar.
For the auto-correlation, the sum is over pairs 
$(\delta_q(\lambda),\delta_{q^\prime}(\lambda^\prime))$
in the $(\rperp,\rpar)$ bin $A$.
The $w_i$ are weights chosen to optimize the measurement by favoring measurements of low-noise and at high-redshift (where the correlation is expected to be highest).
For the cross-correlation, the sum is over pairs of flux-transmission
and quasars,
$(\delta_{q}(\lambda), q^\prime)$.

\item  Calculation of the covariance matrix of the 
auto- and cross-correlations by sub-sampling
(Sections 3.2 and 3.3 of \cite{dmdb2020}).
The method consists of measuring the correlation
functions in different regions of the sky
and deducing the covariances from the variations
between different regions.

\item Fit of the measured auto- and cross-correlations with a 
\lcdm-based model
(Sect. 4 of \cite{dmdb2020})
as described here in Sect. \ref{sec::model}.

\end{enumerate}

\begin{figure}[!htbp]
\centering
\includegraphics[width=\textwidth]{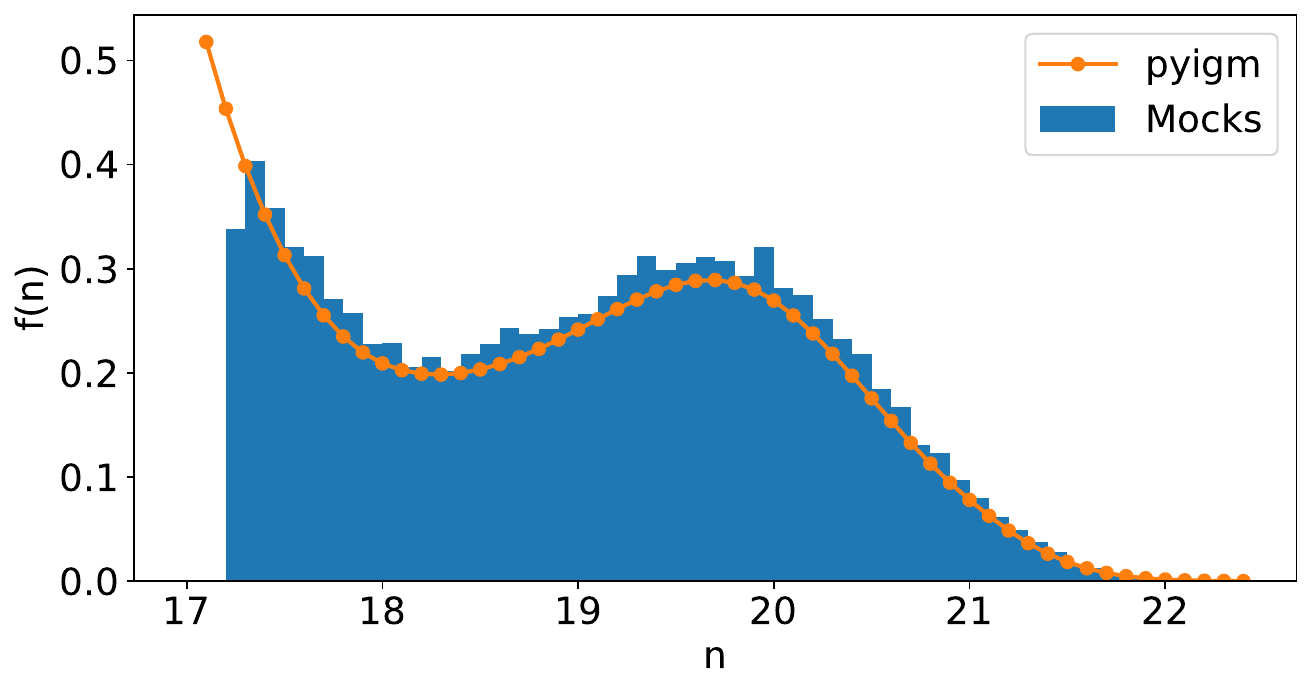}
\caption{The $N_{\text{HI}}$ distribution $f(n=\lognhicm)$ that is used in our analysis. The orange curve shows the distribution of \texttt{pyigm}\citep{prochaska2017pyigm}, and the blue points give the HCD distribution in our mocks.}
\label{fig:fnhi}
\end{figure}

\begin{figure}
\centering
\includegraphics[width=0.7\textwidth]{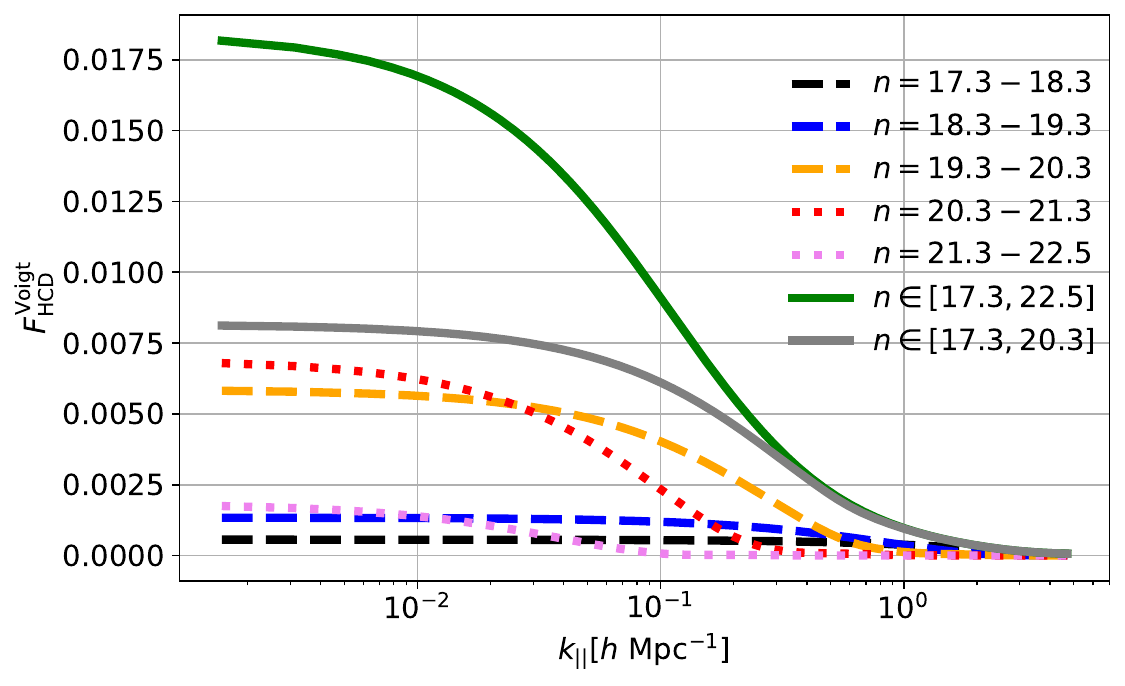}
\caption{The function $\Fhcdvoigt$ (equation \ref{Equa:Fhcd}) based on integration of HCDs with 
the distribution of $\text{N}_{\text{HI}}$ over the ranges as labeled and assuming $\rhohcd=0.00158$\AA$^{-1}$.
The solid green line gives the complete $\Fhcdvoigt$ if no DLAs are masked and the solid gray line the complete $\Fhcdvoigt$ if all DLAs with $\lognhicm>20.3$ are masked.
}
\label{fig:Fvoigt}
\end{figure}

\section{The model for correlations with HCDs}\label{sec::model}

\begin{figure}
\centering
\includegraphics[width=14cm,height=7cm]{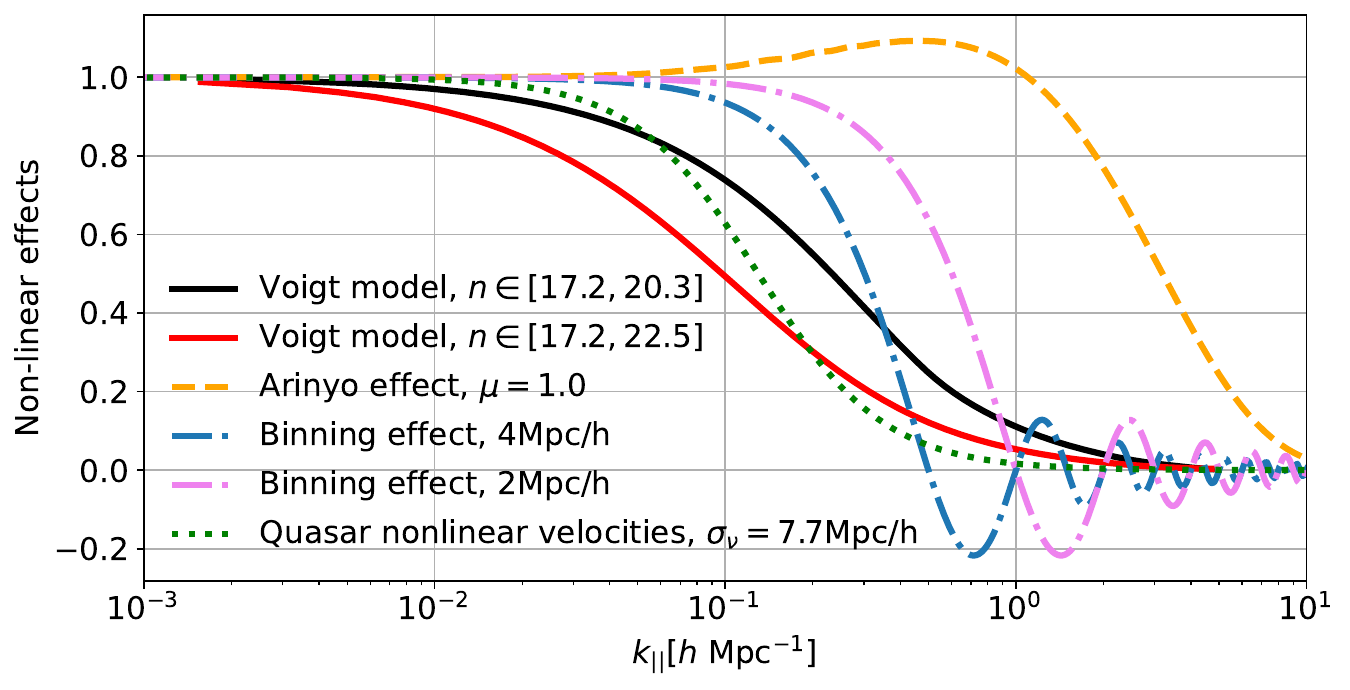}
\caption{Effects of different terms that modify the Ly$\alpha$ forest power spectrum amplitude. The comparison of the $\texttt{Voigt}$ model (Solid lines), the non-linear damping using the Arinyo model (Dashed line), the binning effect (Dotted-dashed lines), and the quasar nonlinear velocities (Dotted line).
}
\label{fig:nonlinear effects}
\end{figure}

Following the procedure used in all BOSS, eBOSS, and DESI analyses,
we model flux-flux and quasar-flux correlations with a biased power spectrum of the underlying \lcdm~matter power spectrum.
For tracers $i$ and $j$ with biases and redshift-space distortion parameters $(b_i,\beta_i)$ and $(b_j,\beta_j)$ the spectrum is
\begin{equation}
    \hat{P}(\kvec) =
    b_{i}b_{j}
    \left( 1+\beta_{i}\mu_{k}^{2}\right)
    \left( 1+\beta_{j}\mu_{k}^{2}\right)
    P_{\rm QL}(\kvec)F_{\rm NL}(\kvec)G(\kvec),
    \label{equation::simple_model_3d_correlation_function}
\end{equation}
where the vector $\kvec = (k_{\parallel},k_{\perp}) = (k, \mu_{k})$ 
of modulus $k$, has components along and across the line-of-sight,
$(k_{\parallel},k_{\perp})$, with
$\mu_{k} = k_{\parallel}/k$.
$P_{\rm QL}$ is the linear \lcdm~power spectrum corrected for bulk flows around the BAO peak.
$F_{\rm NL}$ models the effects on  large-$\vec{k}$ modes of  non-linear growth, Jeans smoothing,  
and line-of-sight broadening due to non-linear velocities and thermal broadening. We use the Arinyo model proposed by
\citep{arinyo2015non} in this study, and compare it with other effects in Figure~\ref{fig:nonlinear effects}.

Finally, $G(\kvec)$ is a damping term that accounts for averaging of the
correlation function in individual $(\rpar,\rperp)$ bins.

For the auto-correlation, the main tracer is the Ly$\alpha$ absorption,
$i=j$, denoted here by the pair $(\blya,\betalya)$.
For the cross-correlation  $i \neq j$, with the second pair 
$(\bquasar,\betaquasar)$.

As described by MW12 and FM12,
the broadening effect due to HCDs can be modeled with an effective scale-dependent bias:
\begin{equation}
\left\{
\begin{array}{ll}
\blya^\prime = \blya  - \bhcd \Fhcdvoigt (\kpar) \\[2mm]
\blya^\prime\betalya^\prime =
\blya \betalya  - \bhcd \betahcd\Fhcdvoigt (\kpar)~~\raisebox{8pt}{,}
\end{array}
\right.
\label{eqn:scale-dependent_bias}
\end{equation}
where $(\blya,\betalya)$ are the bias parameters of the IGM
contribution to forest absorption, and
$(\bhcd,\betahcd)$ are the bias parameters of the halos containing the HCDs.
The function $\Fhcdvoigt$ is
\begin{equation}
    F^{\text{Voigt}}_{\text{HCD}}(k_{||})= \rhohcd \int f(n) W(k_{||},n)dn.\label{Equa:Fhcd}
\end{equation}
Here $\rhohcd$ is the number density of HCDs per unit forest length,
$f(n)$ is the normalized column-density distribution of HCDs, $n=\log N_{HI}$,
and $W(\kpar)$ is the 
Fourier transform of $1-V(\lambda)$ (defined in Eqn.~\ref{eqn:defineV}) where V is a Voigt profile illustrated in Fig. \ref{fig:voigt_profile}.
This form of $\Fhcdvoigt$ assumes $\bhcd$ is independent of $n$.  If not, $\bhcd(n)$ must be included
in the integral.
Possible redshift dependencies of $(\bhcd,\betahcd)$ have also been ignored and we will assume throughout that referring these parameters to their values at the mean redshift is sufficient.
With these restrictions,
a  demonstration of eqns. \ref{eqn:scale-dependent_bias} and \ref{Equa:Fhcd} is given in Appendix A.

The HCD density $\rhohcd f(n)$ is constrained by measurements in the
DLA region \citep{noterdaeme2012column}, $\lognhicm>20.3$, and for
$\lognhicm<17.2$ \citep{Rudie2013}.  
To interpolate between the two extremes,
we use the calculated density of \citep{prochaska2017pyigm} using the package \texttt{pyigm}\footnote{https://github.com/pyigm/pyigm}
shown in Fig. \ref{fig:fnhi}.

Figure \ref{fig:Fvoigt} shows $F^{\text{Voigt}}_{\text{HCD}}$ for different ranges of integration of the $f(n)$ used in the production of the mocks.
The function for the full range of column densities, $17.2<n<22.5$, is the solid green line .
The two major contributors to this function are $20.3<n<21.3$ and $19.3<n<20.3$
with the highest column densities, $21.3<n<22.5$, having a minor importance.
If one masks DLAs with $n>20.3$, the solid grey line is the relevant function with
the major contribution coming from the range $19.3<n<20.3$.


If $f(n)$ is known (as for the mock data), equations \ref{eqn:scale-dependent_bias} and \ref{Equa:Fhcd} show that the power spectrum is determined by
the combinations $\bhcd\rhohcd$ and $\bhcd\betahcd\rhohcd$.
For mock data, the known mean density is 
$\rhohcd(z_{eff}\approx2.29)=0.00158$\AA$^{-1}$,
and by adopting this value fits of the correlation functions return values of $(\bhcd,\betahcd)$.
Since the HCD effect is rather small, the data are not precise enough to determine the two parameters so, following \citep{dmdb2020}, we always place a prior on $\betahcd=0.5\pm0.05$ appropriate for halos of bias $\approx2$ at high redshift.

For real data, $f(n)$ is unknown for $17.2<\lognhicm<20$.
Because of this analyses have used a phenomenological form for $\Fhcd$. The final eBOSS analysis \citep{dmdb2020} 
and the DESI analyses \citep{DESIyr1lyabao,DESI-yr3lyabao} used 
\begin{equation}   -\bhcd \Fhcdvoigt(\kpar)\;\rightarrow\;b^F_{\text{HCD}}
    \exp{(-k_{||}L_{\text{HCD}})}.
\label{Equa:Exp model}
\end{equation}
Here $b_{\text{HCD}}$ is the halo bias of HCDs, and $b^F_{\text{HCD}}$ is the flux bias of HCDs. Since this model uses an exponential function, we call it the $\texttt{Exp}$ model. 
Note that since $\bhcd>0$ and $F_{HCD}^{Voigt}(\kpar=0)>0$, then we must have
$\bfhcd<0$.  In previous BOSS, eBOSS and DESI papers that used the \Expmodel~model, our parameter $\bfhcd$ has
been denoted simply as $\bhcd$.

If the function $\Fhcdvoigt(\kpar)$ is approximately an exponential,
a correspondence between $\bfhcd$ and $\bhcd$ can be found by setting $\kpar=0$ in equ. \ref{Equa:Exp model}:
\begin{equation}
    \bfhcd\;\approx
    -\bhcd\,\rhohcd\,\int\overline{W(\lambda)}d\lambda
\end{equation}
where $\overline{W(\lambda}$ is the mean Voigt profile, $\overline{W}=1-\overline{V}$.
This means that the magnitude of $\bfhcd$ is just $\bhcd$ times the fraction of the forest
pixels with significant absorption due to HCDs.


The power (\ref{equation::simple_model_3d_correlation_function}) contains three functions that suppress high-$\kpar$ modes: $\Fhcd(\kpar)$ modeling the HCD effect; 
$F_{NL}(\kvec)$ due to non-linear IGM effects; and $G(\kvec)$ due to the binning the the correlation function in the estimator.  The cross-correlation for has an additional term due to non-linear quasar velocities, though this effect is not included in the mock data used here. 
Figure ~\ref{fig:nonlinear effects} plots these functions all normalized to unity at $\kpar=0$ though the $\Fhcd$ term affects only the power proportional to $\bhcd$.
The function with the greatest mode suppression is $\Fhcd$ including column densities over the entire range, $17.2<n<22.5$, for which the suppression reaches $\Fhcd=0.5$ at $\kpar\approx0.1\hmpc$.
Including only HCDs with $n<20.3$ (i.e. by masking DLAs) increases this cutoff to $\kpar\approx0.3\hmpc$, comparable with the value for the binning effect with $4\hmpc$ bins.
Note however that in this case $\Fhcd$ has a more gentle decline beginning at lower $\kpar$ values than the sharply cutoff binning effect.
The binning effect can be reduced by using narrower bins, with $2\hmpc$ bins giving a 
cutoff at $\kpar\approx0.7\hmpc$.
The IGM non-linearities of the Arinyo model give a cutoff at $\kpar\approx3\hmpc$.
Finally, the cross-correlation has a cutoff associated with quasar nonlinear velocities, shown on the plot with a smearing parameter $\sigmav=7.7\hmpc$ \citep{dmdb2020}.  This effect is not studied here.

\begin{figure}[!htbp]
\centering
\includegraphics[width=0.48\textwidth]{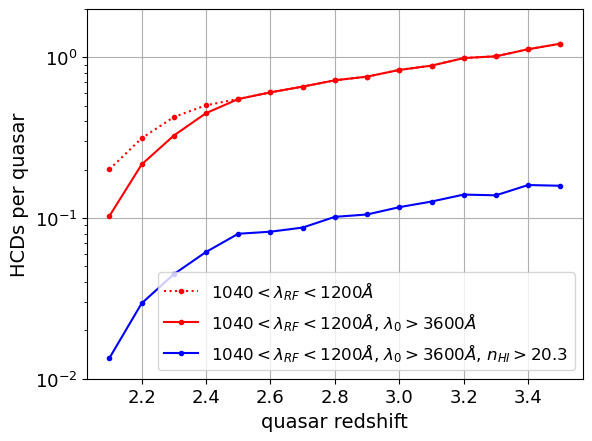}
\includegraphics[width=0.46\textwidth]{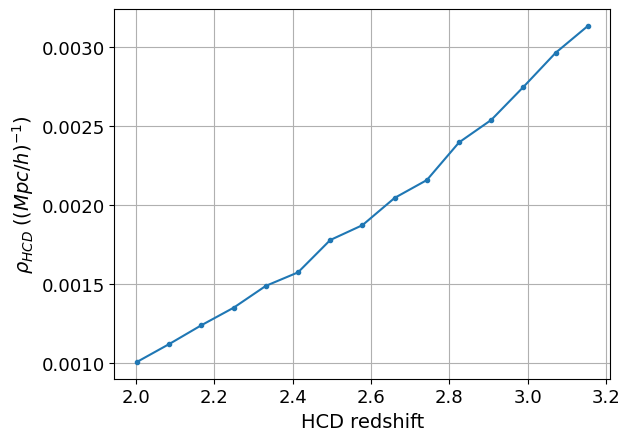}
\caption{Statistics of HCDs as a function of redshift in the mock dataset: the number of HCDs per forest as a function of quasar redshift for the mock data (plot on the left); the density of HCDs as a function of redshift for the mock data (plot on the right).}
\label{fig:rhohcd}

\label{fig:hcd_combined}
\end{figure}

\begin{figure}
\centering
\includegraphics[width=\textwidth]{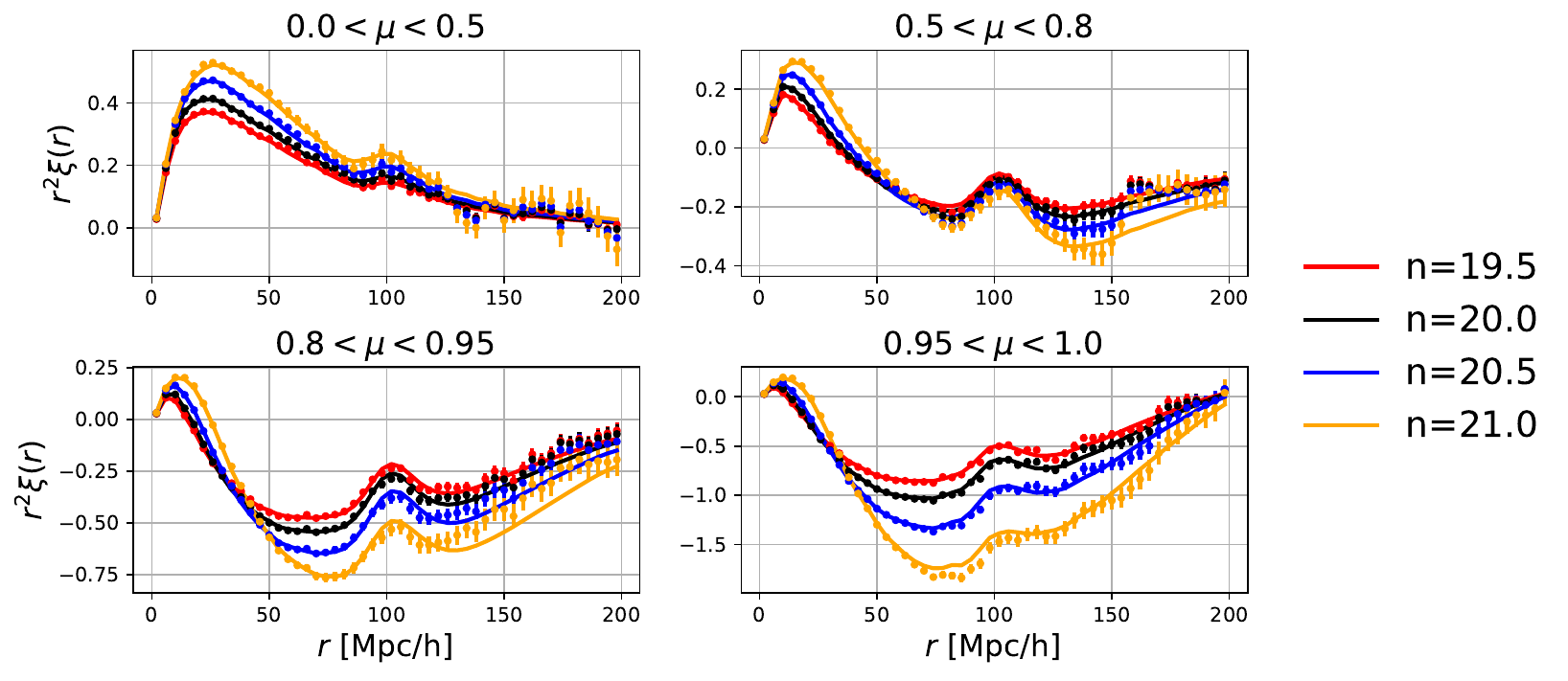}
\newline
\newline
\newline
\includegraphics[width=\textwidth]{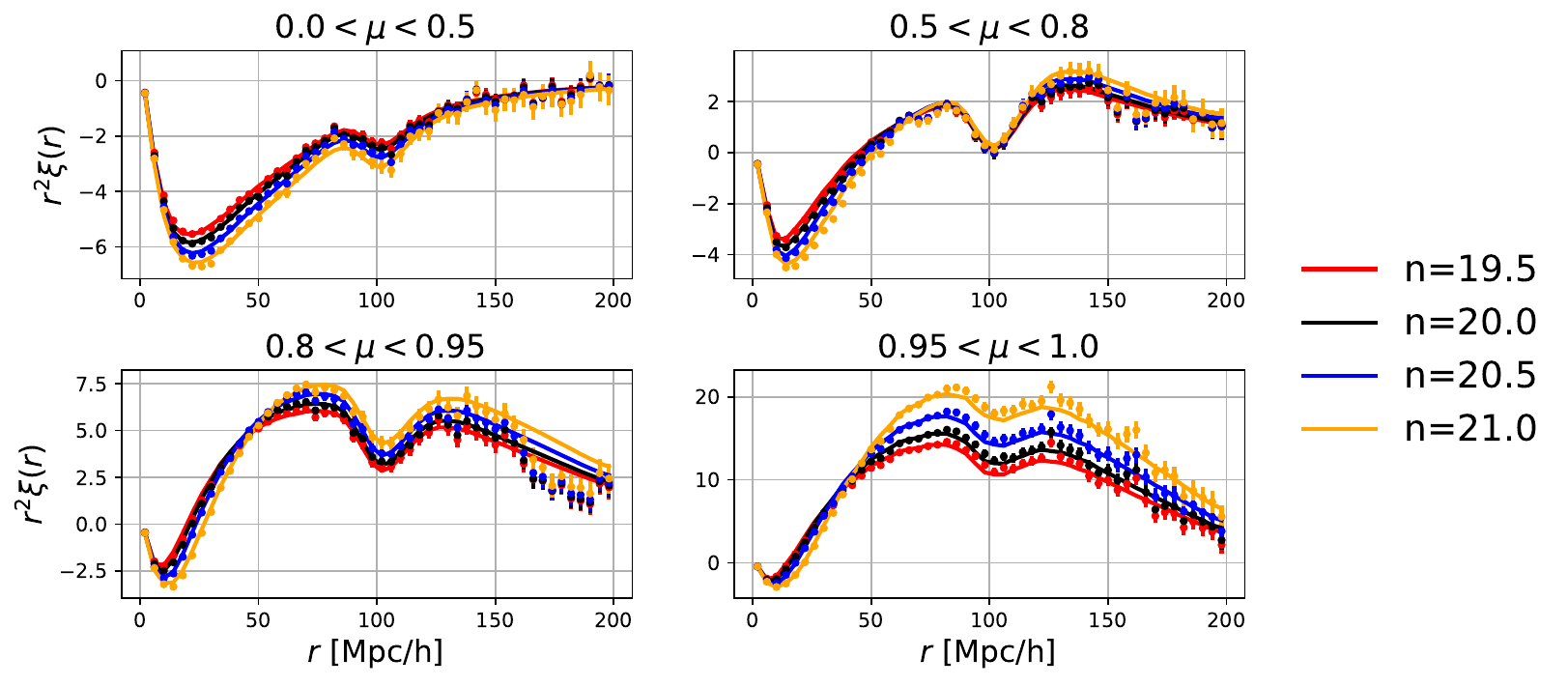}
\caption{The Ly$\alpha$ auto-correlation function (top four panels) and Ly$\alpha$-quasar cross-correlation (bottom four panels), for mock data. The points of different colors give the
measured correlation for mocks with HCDs with unique column densities $\lognhicm=19.5,20,20.5,21$, and the curves give the best fit models using the Voigt model, in four wedges of $|\mu|=|\frac{r_{||}}{r}|$. The fitted range is chosen as $r\in[20,180] h^{-1}\text{Mpc}$.
}

\label{fig:auto cross correlation function - 1}
\end{figure}

\section{Tests with mocks}\label{sec:testswithmocks}

We used the mock data sets described in \citet{Etourneau2024}
which  have been used to verify that the DESI analysis accurately measures the BAO parameters of the \lya~forest auto-correlation and the quasar-forest cross-correlation \citep{DESIyr1lyabao,DESI-yr3lyabao}.
Here, we
test the Voigt-based physical model and  the phenomenological \Expmodel~model used to model HCD effects.
In particular, we verify that the fits with the \Voimodel~model recover the known HCD bias, $\bhcd$.
A secondary aim is to study degeneracies between the HCD bias  and the IGM bias parameters, $\blya$ and $\betalya$.


The version of the mocks for this study use Gaussian-random fields produced on a grid covering the eBOSS survey to define the linear density, velocity and velocity-gradient fields.
These Gaussian fields were then used to generate data sets containing quasars and associated \lya~forests and HCDs as follows:

\begin{itemize}
    \item \textbf{Quasars} were placed using a lognormal density field derived from the Gaussian random field, so that their clustering follows the Kaiser power spectrum with desired bias and RSD parameters.
    
    \item \textbf{IGM absorption}
    was generated from  Gaussian random fields using
    the Fluctuating Gunn-Perterson Approximation (FGPA) \citep{FGPHreference}.
    
    \item \textbf{High-column-density systems (HCDs)} were placed at density peaks with a threshold defined by equation A2 of \citep{font2012effect}
so as to produce an HCD bias of $\bhcd=2$ and $\betahcd=0.5$. These values are motivated by measurments of the HCD-forest cross-correlation by  \citet{perez2023cross}.
The details of the insertion of HCDs and other astrophysical effects is described in  \citet{Herrera_2024}. The number density and redshift evolution are matched to the distribution from \texttt{pyigm}, calibrated to observations over $17.2 < \log N_{\rm HI} < 22.5$ and $2.0<z_{\rm{HCD}}<3.2$. 

\end{itemize}


The mocks are generated to cover the SDSS-IV survey footprint and quasar density, with at the end, a catalog of $261854$ quasars.
The number of HCDs per quasar and 
the density of HCDs are shown in Fig. \ref{fig:rhohcd}. 
For a typical eBOSS quasar of redshift 2.5, there are $\approx0.6$ HCDs per quasar corresponding to density of $\approx0.0016(\hmpc)^{-1}$.





In addition to mocks produced with the \texttt{pyigm} distribution of $\lognhi$,
in order to facilitate the testing of the HCD correlation model,
special mock sets were generated with all HCDs given a unique value of $\lognhi$.
Sets were produced with the values $n=19.5,20.0,20.5,$ and $21$. For these mocks, correlation functions were computed with no DLAs masked.
For mocks with the \texttt{pyigm} distribution $f(n)$, fits were performed both with and without
masking DLAs, the former corresponding to $n\in[17.2,20.3]$ and the latter to $n\in[17.2,22.5]$.

The complete auto- and cross-correlation functions for stacks of 10 Ly$\alpha$ mocks are shown in Figure~\ref{fig:auto cross correlation function - 1} for the mock sets with unique values of $n$.

For the fits of mock data, we have $6$ free parameters $\{\alpha_{||},\alpha_{\perp},\biasetalya,\betalya,\bhcd,\betahcd\}$ in the fitting of the auto-correlation function, where $\biasetalya=\blya\times \betalya$.
As in \citet{dmdb2020} we find that the fits do not constrain well the parameter $\betahcd$, so we use a Gaussian prior $\betahcd=0.5\pm0.09$.

For the cross-correlation, we have one more free parameter, $\Delta\rpar$, that models any systematic shift in quasar redshifts relative to that of the $\lambda$ grid. 
The parameters $b_{\text{QSO}}$ and $\beta_{\text{QSO}}$ are fixed because they are not significantly constrained by the cross-only fits and we do not perform joint auto-cross fits.

\begin{table*}
\begin{center}
\scalebox{0.85}{\begin{tabular}{|c|c|c|c|c|}
\hline
&$b_{{\rm{{HCD}}}}$&$\beta_{{\rm{{HCD}}}}$&$\biasetalya$&$\beta_{{\rm{{Ly}}\alpha}}$\\ 
$\rm{Ly}\alpha\times\rm{Ly}\alpha$ && & &  \\
$n=19.5$&1.95$\pm$0.292&0.48$\pm$0.09&-0.206$\pm$0.002&1.61$\pm$0.04\\ 
$n=20.0$&2.13$\pm$0.157&0.48$\pm$0.08&-0.204$\pm$0.003&1.60$\pm$0.05\\ 
$n=20.5$&2.2$\pm$0.092&0.5$\pm$0.07&-0.201$\pm$0.003&1.57$\pm$0.06\\ 
$n=21.0$&1.92$\pm$0.052&0.75$\pm$0.05&-0.187$\pm$0.003&1.33$\pm$0.05\\ 
$n\in[17.2,20.3]$&1.93$\pm$0.346&0.48$\pm$0.09&-0.206$\pm$0.001&1.64$\pm$0.04\\ 
$n\in[17.2,22.5]$&1.82$\pm$0.146&0.48$\pm$0.08&-0.205$\pm$0.002&1.64$\pm$0.04\\ 
no HCDs & - & - & -0.2077$\pm$0.0005 &1.687$\pm$0.008\\ 
\hline
$\rm{Ly}\alpha\times\rm{QSO}$ && & &  \\
$n=19.5$&2.3$\pm$0.488&0.51$\pm$0.09&-0.194$\pm$0.004&1.67$\pm$0.08\\ 
$n=20.0$&2.64$\pm$0.427&0.53$\pm$0.09&-0.191$\pm$0.006&1.78$\pm$0.13\\ 
$n=20.5$&2.59$\pm$0.275&0.56$\pm$0.08&-0.19$\pm$0.007&1.89$\pm$0.18\\ 
$n=21.0$&2.12$\pm$0.138&0.63$\pm$0.08&-0.191$\pm$0.007&1.67$\pm$0.14\\ 
$n\in[17.2,20.3]$&2.2$\pm$0.49&0.51$\pm$0.09&-0.193$\pm$0.003&1.63$\pm$0.05\\ 
$n\in[17.2,22.5]$ & 2.46$\pm$0.387 & 0.53$\pm$0.09 & -0.191$\pm$0.004 & 1.78$\pm$0.11\\ 
no HCDs & - & - & -0.190$\pm$0.001 & 1.564$\pm$0.015\\
\hline
\end{tabular}}
\caption{
Best-fit Voigt model values of the four bias parameters for stacks of 10 mocks.
The parameters are shown for the different distributions of $n=\lognhi$ described in the first column.  The complete sets of best-fit parameters are shown in Tables \ref{Tab:voigt_fits_unique_n} and \ref{Tab:fits_realistic_n}.
Also shown are the fits for $(\biasetalya,\betalya)$ for mocks with no HCDs (table 3 of \citep{Etourneau2024}). }
\label{tab:biasparameters_voigt}
\end{center}
\end{table*}

\begin{figure}
\centering
\includegraphics[width=0.7\textwidth]{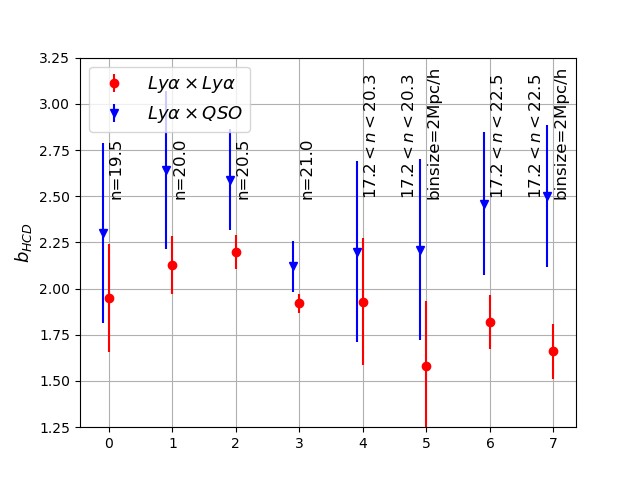}
\caption{Measurements of $b_\text{HCD}$
using the Voigt model for mocks with different distributions of $\lognhi$. 
}
\label{fig:bhcd_voigt}
\end{figure}

\begin{figure}
\centering
\includegraphics[width=0.7\textwidth]{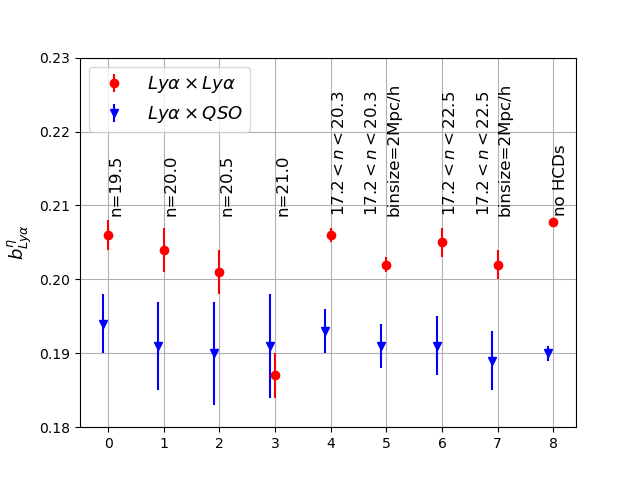}\\
\includegraphics[width=0.7\textwidth]{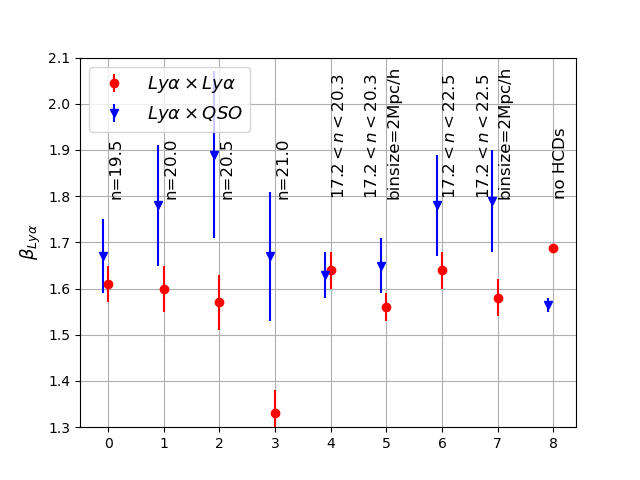}
\caption{Measurements of $\biasetalya$ and $\betalya$
using the Voigt model for mocks with different distributions of $\lognhi$. 
}
\label{fig:blya_voigt}
\end{figure}

\subsection{Fits with the Voigt model}

The best-fit bias parameters, $(\bhcd,\betahcd,\biasetalya,\betalya)$, for stacks of 10 mocks for the Voigt model are shown in Table \ref{tab:biasparameters_voigt} 
and plotted in Figs. \ref{fig:bhcd_voigt} and \ref{fig:blya_voigt}.
The best-fit values for all parameters are listed in Tables~\ref{Tab:voigt_fits_unique_n} and \ref{Tab:fits_realistic_n}. 
 The most important result is that the values of $\bhcd$ are generally within two standard deviations of the expected value, $\bhcd=2$, thus accomplishing the primary aim of this study.

With the exception of the $n=21$ case, the  values for $(\biasetalya,\betalya)$ are stable with respect to changes in the distribution of $\lognhi$ and within $\approx5\%$ of the value obtained with mocks containing no HCDs \citep{Etourneau2024}.
As observed in \citep{Etourneau2024}, there is a $\approx10\%$ difference in the $\biasetalya$ values obtained for the auto- and cross-correlations and a $\approx7\%$ differnce for $\betalya$.
In all cases the BAO paramters $(\aperp,\apar)$ are consistent with the expected values of unity (Tables~\ref{Tab:voigt_fits_unique_n} and \ref{Tab:fits_realistic_n}).



The fits of the mocks with $n=21$ have poor values of $\chi^2$ (Table~\ref{Tab:voigt_fits_unique_n}) and anomalous values of $(\betahcd,\biasetalya,\betalya)$.
This is most likely due to the fact that for such large values of $n$ the width of the DLA becomes comparable to the width of the forest.  This interferes with the fit of each forest continuum, described in step 2 of Sect. \ref{sec::analprocedure}, that is necessary to determine the forest continuum about which fluctuations are measured.
Because such high-$n$ DLAs are rare and easily masked, we leave this problem for future study.

\begin{figure}
\centering
\includegraphics[width=14cm,height=14cm]{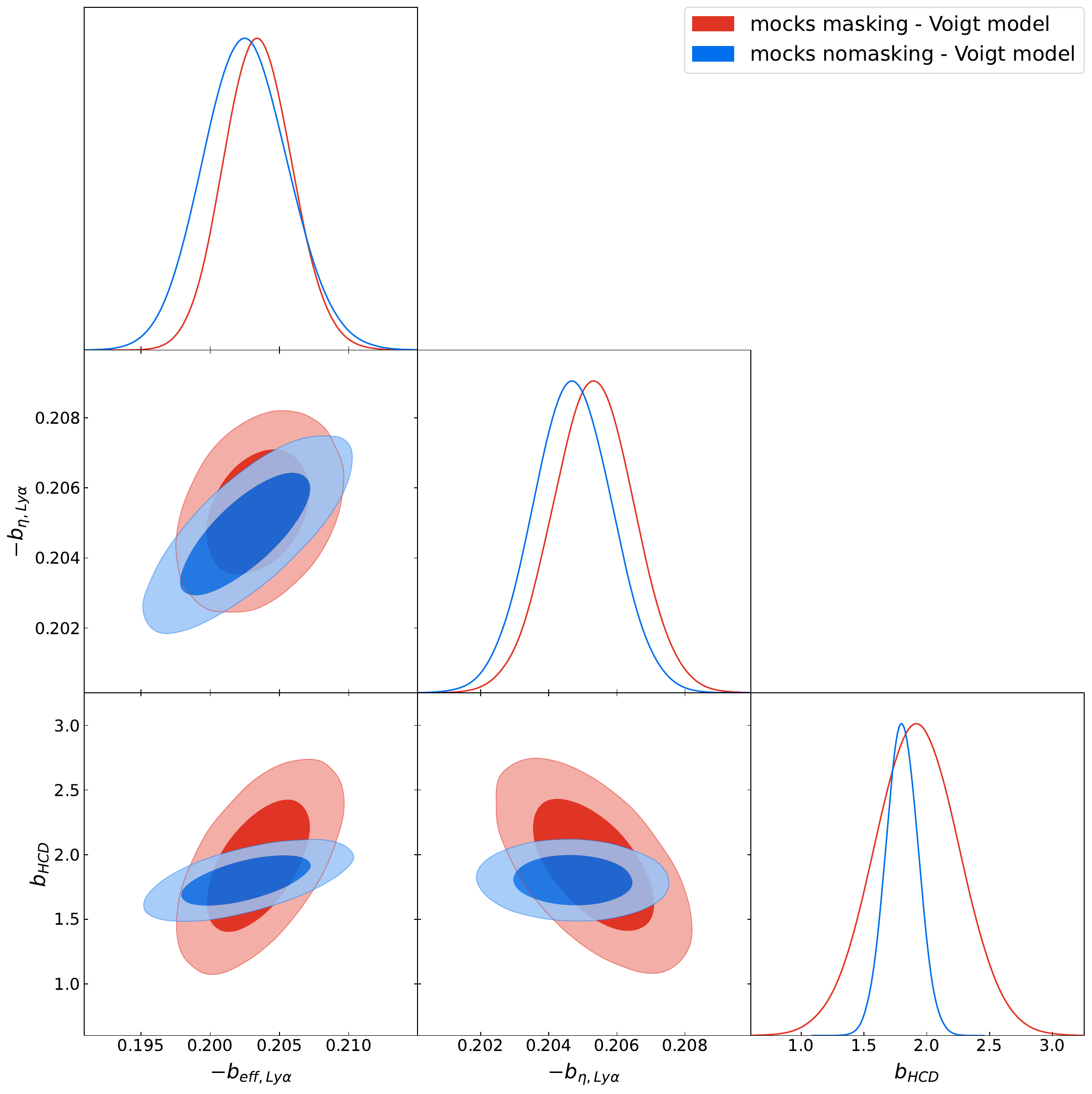}
\caption{
Correlated constraints for the Ly$\alpha$ parameters constraints \{$b_{\text{eff},\text{LY}\alpha}$,$b_{\eta,\text{LY}\alpha}$,$b_{\text{HCD}}$\} using the $\texttt{Voigt}$ model fits of the auto-correlation. 
}
\label{fig:triangle-voigt}
\end{figure}

Figure~\ref{fig:triangle-voigt} shows the correlated constraints on \{$b_{\text{eff},\text{LY}\alpha}$,$b_{\eta,\text{LY}\alpha}$,$b_{\text{HCD}}$\} for the $\texttt{Voigt}$ model fits of the auto-correlations. 
We use the bias determining the monopole $b_{\text{eff},\text{LY}\alpha}=b_{\text{LY}\alpha}(1+\frac{2}{3}\beta_{\text{LY}\alpha}+\frac{1}{5}\beta_{\text{LY}\alpha}^2)^{1/2}$ instead of $\blya$ since it is less correlated with $\betalya$ \citep{Etourneau2024}.
We do not show $\betahcd$ since it is mostly determined by priors.
For nomasking, there is a non-negligible anti-correlation between $\bhcd$ and $|\biasetalya|$, corresponding to a fixed total bias.

\subsection{Comparison with the Exp model}

Fits were performed the \Expmodel~model on the mock data with  the \texttt{pyigm} distribution $f(n)$ with both masking and no-masking. Compared to the \Voimodel~model, the \Expmodel~model has one additional free parameter $L_{\text{HCD}}$.

Table \ref{tab:voigt_exp_parameters} shows the best-fit bias parameters for the  \Voimodel~and  \Expmodel~models.
The complete set of  fit parameters is shown in Table~\ref{Tab:fits_realistic_n} 
where we see that both models give acceptable values of $\chi^2$ and the difference between them is not sufficient to establish the superiority
of one over the other.

\begin{figure}
\centering
\includegraphics[width=\textwidth]{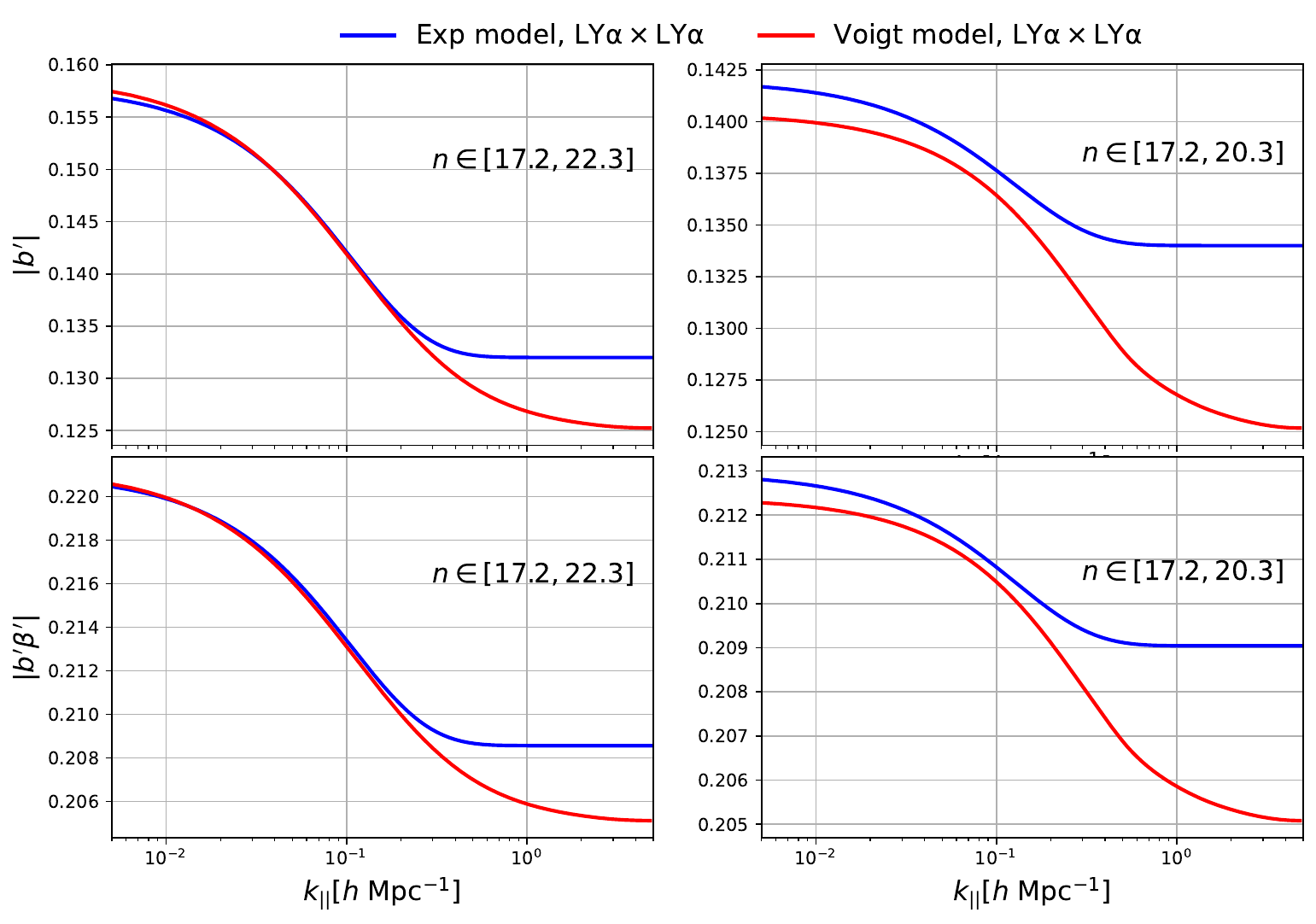}
\caption{
The scale-dependent biases $b^\prime \beta^\prime$ and $b^\prime$ from eqn.~\ref{eqn:scale-dependent_bias} as a function of $\kpar$
for the \Voimodel~model (blue) and \textbf{from eqn.~\ref{Equa:Exp model} for} the \Expmodel~model (red).
The biases are shown for the full range of $\nhi$  (left) and for masked DLAs (right)
as determined by fitting the auto-correation function.
}
\label{fig:comparison model - 1}
\end{figure}

The functions $b^\prime(\kpar)$ and $b^\prime(\kpar)\beta^\prime(\kpar)$
are shown in Figure~\ref{fig:comparison model - 1}.
The \Voimodel~and \Expmodel~ model functions are very similar except
at large $\kpar$ where the other damping components (Fig. \ref{fig:nonlinear effects}) are important.

\begin{figure}
\centering
\includegraphics[width=14cm,height=14cm]{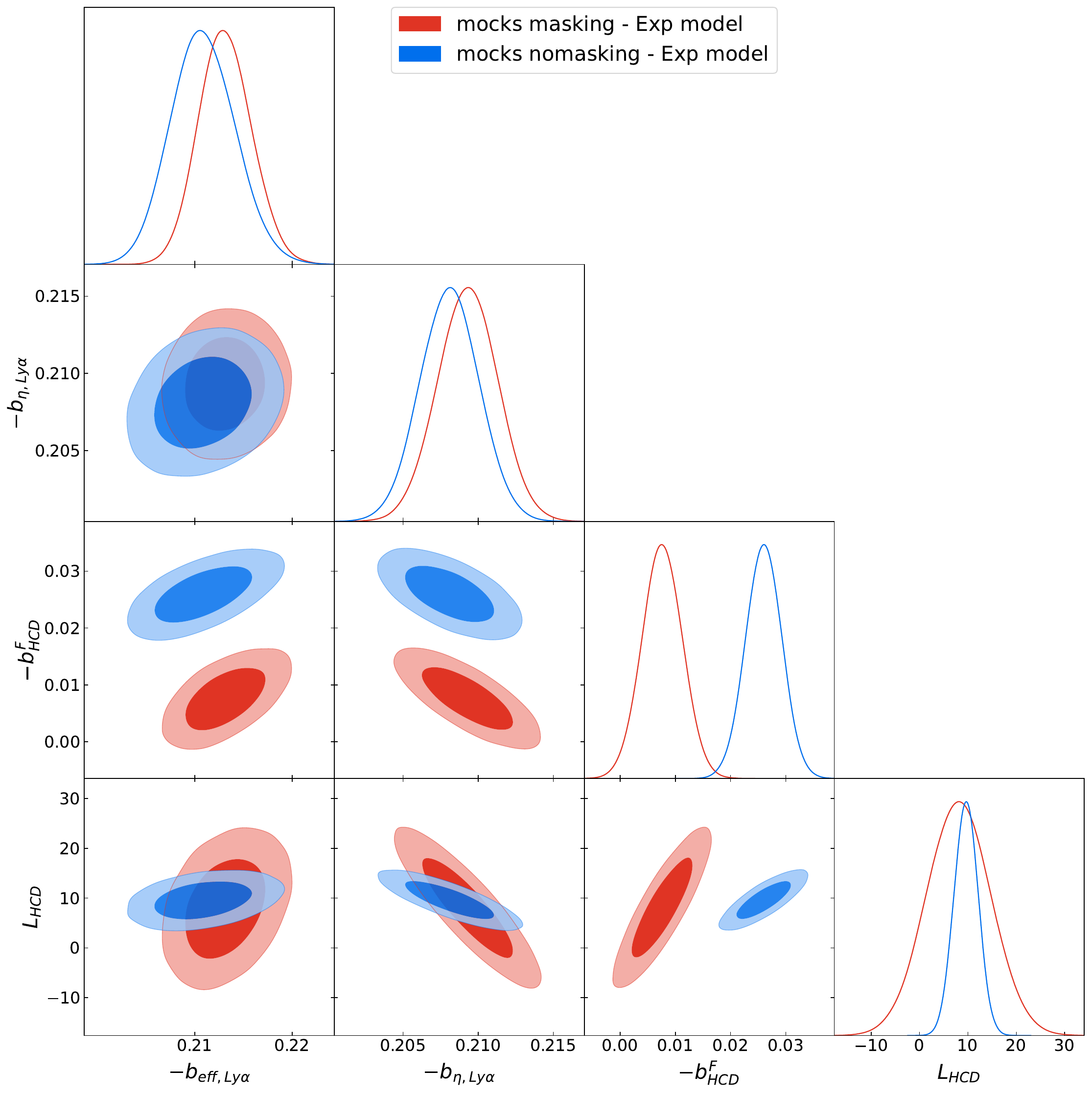}
\caption{For the \Expmodel~model, the
correlated constraints for the Ly$\alpha$ parameters  \{$b_{\text{eff},\text{LY}\alpha}$,$b_{\eta,\text{LY}\alpha}$,$b^F_{\text{HCD}}$,$L_{\text{HCD}}$\}.
The one and two standard deviation contours are calculated assuming a Gaussian likelihood so
for the marginally measured parameter, $\Lhcd$, the two-standard deviation curves are
not realistic. 
}
\label{fig:triangle-exp}
\end{figure}

Figure~\ref{fig:triangle-exp} shows the correlated constraints  of the Ly$\alpha$ bias parameters  \{$b_{\text{eff},\text{LY}\alpha}$, $b_{\eta,\text{LY}\alpha}$, $b^F_{\text{HCD}}$, $L_{\text{HCD}}$\} for the \Expmodel~model.  
For masked mocks, 
the constraints on the two HCD parameters are weak:
$\bfhcd=-0.008\pm0.004$ and $\Lhcd=7.8\pm6.6$.  
The one parameter of the \Voimodel~model is determined with better precision:
$\bhcd=1.93\pm0.35$.
For the unmasked mocks, both models give significant non-zero values for
the HCD bias.

The values of  IGM parameters $(\biasetalya,\betalya)$ are comparable for the masking and no-masking cases.

\begin{figure}
\centering
\includegraphics[width=14cm,height=14cm]{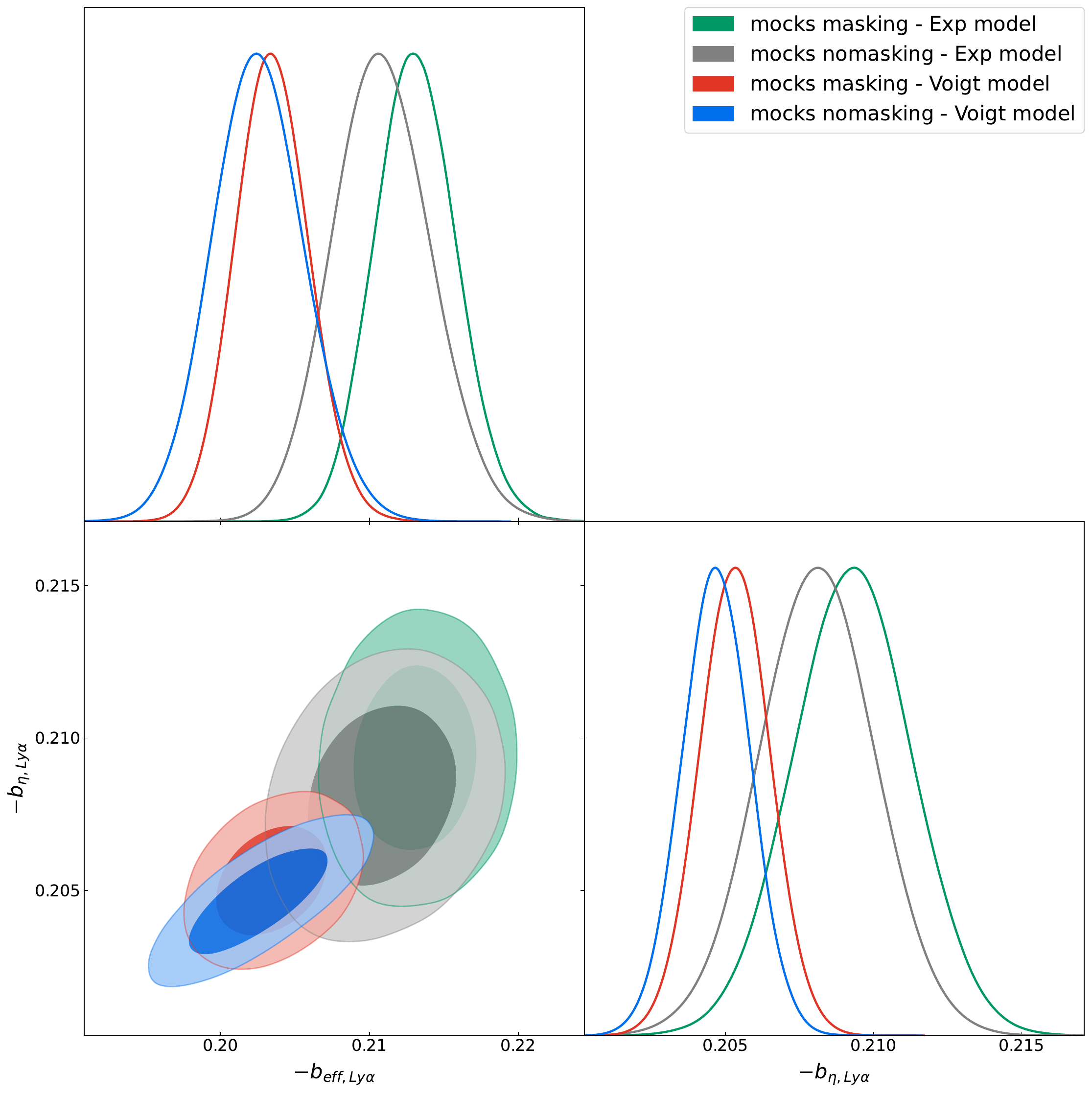}
\caption{
The comparison of the constraints on the Ly$\alpha$ IGM bias parameters \{$b_{\text{eff},\text{LY}\alpha}$,$b_{\eta,\text{LY}\alpha}$\} between the $\texttt{Voigt}$ model and the $\texttt{Exp}$ model fits of the auto-correlation.}
\label{fig:triangle-exp+voigt}
\end{figure}

A comparison of the \Voimodel~ and \Expmodel~models for the auto-correlation fits of the IGM bias parameters \{|$b_{\text{eff},\text{LY}\alpha}$|,$|b_{\eta,\text{LY}\alpha}|$\} is shown
in Figure~\ref{fig:triangle-exp+voigt}.
The \Voimodel~model gives tighter constraints on both IGM parameters.

\begin{table*}
\begin{center}
\scalebox{0.85}{\begin{tabular}{|c|c|c|c|c|c|c|}

\hline
&$b_{{\rm{{HCD}}}}$&$b^F_{{\rm{{HCD}}}}$&$L_{\rm{HCD}}$&$\beta_{{\rm{{HCD}}}}$&$b_{{\eta,\rm{{Ly}}\alpha}}$&$\beta_{{\rm{{Ly}}\alpha}}$\\ 
& \Voimodel~model & \Expmodel~model & \Expmodel~model & & & \\
$\rm{Ly}\alpha\times\rm{Ly}\alpha\quad\rm{ Mocks}$ && & &&&  \\
$\rm{binsize=4}h^{-1}\rm{Mpc}$ && & &&&  \\
$n\in[17.2,20.3]$&1.93$\pm$0.346&&&0.48$\pm$0.09&-0.206$\pm$0.001&1.64$\pm$0.04\\ 
&&-0.008$\pm$0.004&7.86$\pm$6.64&0.49$\pm$0.09&-0.209$\pm$0.002&1.56$\pm$0.04\\ 
$n\in[17.2,22.5]$&1.82$\pm$0.146&&&0.48$\pm$0.08&-0.205$\pm$0.002&1.64$\pm$0.04\\ 
&&-0.026$\pm$0.004&9.5$\pm$2.56&0.48$\pm$0.09&-0.208$\pm$0.002&1.58$\pm$0.05\\ 
\hline
&$b_{{\rm{{HCD}}}}$&$b^F_{{\rm{{HCD}}}}$&$L_{\rm{HCD}}$&$\beta_{{\rm{{HCD}}}}$&$b_{{\eta,\rm{{Ly}}\alpha}}$&$\beta_{{\rm{{Ly}}\alpha}}$\\ 
$\rm{Ly}\alpha\times\rm{Ly}\alpha\quad\rm{ Mocks}$ && & &&&  \\
$\rm{binsize=2}h^{-1}\rm{Mpc}$ && & &&&  \\
$n\in[17.2,20.3]$&1.58$\pm$0.355&&&0.46$\pm$0.09&-0.202$\pm$0.001&1.56$\pm$0.03\\ 
&&-0.004$\pm$0.002&15.1$\pm$11.4&0.48$\pm$0.09&-0.206$\pm$0.001&1.48$\pm$0.02\\ 
$n\in[17.2,22.5]$&1.66$\pm$0.149&&&0.47$\pm$0.09&-0.202$\pm$0.002&1.58$\pm$0.04\\ 
&&-0.024$\pm$0.004&9.12$\pm$2.79&0.47$\pm$0.09&-0.205$\pm$0.002&1.53$\pm$0.05\\ 
\hline
\hline
&$b_{{\rm{{HCD}}}}$&$b^F_{{\rm{{HCD}}}}$&$L_{\rm{HCD}}$&$\beta_{{\rm{{HCD}}}}$&$b_{{\eta,\rm{{Ly}}\alpha}}$&$\beta_{{\rm{{Ly}}\alpha}}$\\ 
$\rm{Ly}\alpha\times\rm{Ly}\alpha\quad\rm{ DR16}$ && & &&&  \\
$\rm{binsize=4}h^{-1}\rm{Mpc}$ && & &&&  \\
$n\in[17.2,20.3]$&7.3$\pm$0.611&&&0.67$\pm$0.08&-0.179$\pm$0.004&1.71$\pm$0.11\\ 
&&-0.105$\pm$0.022&2.28$\pm$0.63&0.53$\pm$0.08&-0.175$\pm$0.013&3.23$\pm$1.26\\ 
$n\in[17.2,22.5]$&4.79$\pm$0.326&&&0.67$\pm$0.08&-0.189$\pm$0.005&1.84$\pm$0.14\\ 
&&-0.139$\pm$0.02&2.59$\pm$0.52&0.51$\pm$0.08&-0.173$\pm$0.013&5.25$\pm$3.29\\ 
\hline
\hline
\hline
$\rm{Ly}\alpha\times\rm{QSO}\quad\rm{ Mocks}$ && &&& &  \\
$\rm{binsize=4}h^{-1}\rm{Mpc}$ && & &&&  \\
$n\in[17.2,20.3]$&2.2$\pm$0.49&&&0.51$\pm$0.09&-0.193$\pm$0.003&1.63$\pm$0.05\\ 
&&-0.077$\pm$0.024&2.56$\pm$1.22&0.51$\pm$0.07&-0.169$\pm$0.015&2.83$\pm$0.92\\ 
$n\in[17.2,22.5]$&2.46$\pm$0.387&&&0.53$\pm$0.09&-0.191$\pm$0.004&1.78$\pm$0.11\\ 
&&-0.082$\pm$0.06&4.97$\pm$4.64&0.52$\pm$0.09&-0.176$\pm$0.036&2.83$\pm$1.74\\ 
\hline
$\rm{Ly}\alpha\times\rm{QSO}\quad\rm{ Mocks}$ && &&& &  \\
$\rm{binsize=2}h^{-1}\rm{Mpc}$ && & &&&  \\
$n\in[17.2,20.3]$&2.21$\pm$0.49&&&0.51$\pm$0.09&-0.191$\pm$0.003&1.65$\pm$0.06\\ 
&&-0.016$\pm$0.005&32.3$\pm$33.3&0.51$\pm$0.09&-0.2$\pm$0.003&1.6$\pm$0.08\\ 
$n\in[17.2,22.5]$&2.5$\pm$0.385&&&0.53$\pm$0.09&-0.189$\pm$0.004&1.79$\pm$0.11\\ 
&&-0.074$\pm$0.012&5.87$\pm$1.69&0.52$\pm$0.08&-0.18$\pm$0.009&1.6$\pm$0.31\\ 
\hline
\hline
$\rm{Ly}\alpha\times\rm{QSO}\quad\rm{ DR16}$ && &&& &  \\
$\rm{binsize=4}h^{-1}\rm{Mpc}$ && & &&&  \\
$n\in[17.2,20.3]$&3.78$\pm$1.92&&&0.51$\pm$0.09&-0.237$\pm$0.014&1.91$\pm$0.21\\ 
&&-0.034$\pm$0.024&0.95$\pm$2.87&0.52$\pm$0.09&-0.228$\pm$0.016&1.91$\pm$0.33\\ 
$n\in[17.2,22.5]$&-0.424$\pm$1.4&&&0.5$\pm$0.09&-0.27$\pm$0.018&1.56$\pm$0.16\\ 
&&-0.047$\pm$0.027&-0.01$\pm$1.66&0.51$\pm$0.09&-0.231$\pm$0.019&1.91$\pm$0.34\\ 
\hline

\end{tabular}}
\caption{
Best-fit \Voimodel~and \Expmodel~model values of the bias parameters for stacks of 10 mocks and for the eBOSS dr16 data.
For the \Voimodel~model, the HCD bias parameters are $(\bhcd,\betahcd)$ and for the \Expmodel~model the bias parameters are $(\bfhcd,\Lhcd,\betahcd)$.
The parameters are shown for the two \texttt{pyigm} distributions of $n=\lognhi$ described in the first column and for two bins sizes.  The complete sets of best-fit parameters are shown in Tables \ref{Tab:voigt_fits_unique_n} and \ref{Tab:fits_realistic_n}.
}
\label{tab:voigt_exp_parameters}
\end{center}
\end{table*}

\begin{figure}
\centering
\includegraphics[width=0.48\textwidth]{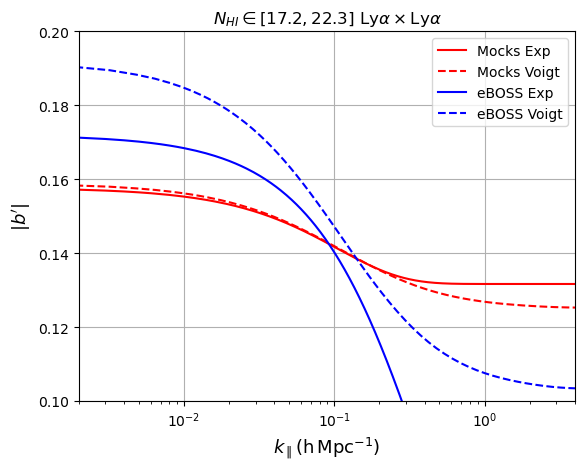}
\includegraphics[width=0.48\textwidth]{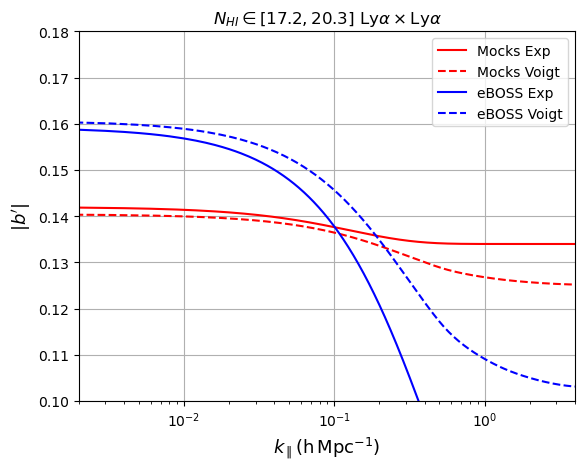}
\caption{
The scale-dependent bias $b^\prime$ from eqn.~\ref{eqn:scale-dependent_bias} as a function of $\kpar$
for the mock data (red) and the eBOSS data (blue).
The dashed lines are for the \Voimodel~model  and the solid lines for  the \Expmodel~model.
The bias is shown for the full range of $\nhi$  (left) and for masked DLAs (right)
as determined by fitting the auto-correation function.
}
\label{fig:eboss_biass}
\end{figure}

\section{Fits with eBOSS data}
\label{sec:eBOSS}

Application of the \Voimodel~model to real data involves two difficulties.
First, the fits require knowledge of the density of HCDs, $\rho f(n)$.  While this is known for the mock data, it can only be estimated for the data using the known distribution of column densities of DLAs and extended to $n<20.3$ with the \texttt{pyigm} distribtion.
Second,
if the identified DLAs are masked, it is necessary to understand the detection efficiency of DLAs and model the effects of unmasked DLAs, as well as the impact of masking incorrectly non-DLA structures.

Two strategies can address these two problems.  First, one could simply not mask any identified
DLAs and model  the full HCD distribution using the \texttt{pyigm} distribution normalized so as
to reproduce the distribution of DLAs measured in high signal-to-noise forests.  An alternative
strategy would be to mask only DLAs found in  high-S/N forests and model the correlations with  with two $\Fhcd$'s, one for the masked high-S/N forests and one for the unmasked low-S/N forests.
This strategy has not yet been implemented, so here we adopt the strategy of masking  identified DLAs and modeling the correlations with an \Expmodel~model, as in \citep{dmdb2020}, or with a \Voimodel~model accounting for HCDs with $n<20.3$.

Compared to the mock data used in the previous section,
fits with real data need $7$ additional free parameters to account for effects not included in the mocks: five to  characterize the different metal absorption lines overlapping with the Ly$\alpha$ absorption in real data, and two parameters that model the sky subtraction residuals \cite{dmdb2020}. 

These parameters are not expected to be degenerate with the HCD parameters:
the major metal parameters are well determined by peaks in the correlations at small $\rperp$ and the sky subtraction parameter is determined by excess correlations at $\rpar=0$.
For the cross correlation, we also fit a parameter, $\sigma_v$ taking into account the quasar velocity distribution parameter.
As indicated by Fig. \ref{fig:nonlinear effects}, this gives a lower cutoff in $\kpar$ if the DLAs are masked, making the HCD effect uncertain.

The fits in the final eBOSS analysis \citep{dmdb2020} used the \Expmodel~model but assumed $\Lhcd=10\hmpc$.  As with the mock analysis, here we take $\Lhcd$ as a free parameter.
Fits were performed both by masking identified DLAs and without masking. 
The parameter $\betahcd$ is mainly determined by the input Gaussian prior $\betahcd=0.5\pm0.09$.
For the Voigt fits, $\rho f(n)$ was taken to be that of \texttt{pyigm}. 

Table \ref{tab:voigt_exp_parameters} gives the values for the bias parameters, and Table~\ref{Tab:fits_eboss} shows the results for all the fitted parameters.

The derived scale-dependent bias (eqn. \ref{eqn:scale-dependent_bias}) is plotted in Fig. \ref{fig:eboss_biass} and compared with that of the mocks.
For the auto-correlation, the derived values of $\bhcd$ and $\bfhcd$ are significantly larger than those found for the mocks.
With no masking we find $\bhcd=7.3\pm0.6$ for DR16 data compared to $1.93\pm0.35$ for the mocks.
With masking we find $\bhcd=4.8\pm0.63$ for DR16 data compared to $1.82\pm0.15$ for the mocks.

These large values of $\bhcd$ are the reason for the much stronger mode-suppression at high $\kpar$ seen for the eBOSS data in Fig. \ref{fig:eboss_biass}

The value of $\Lhcd$ determined for the \Expmodel~model is very low (less than the $\rpar$ bin size) and does not change when there is no masking.
In Fig. \ref{fig:eboss_biass} this causes the cutoff  for the \Expmodel~model to be at lower values of $\kpar$ than the cutoff for the \Voimodel~model where the cutoff $\kpar$ is determined by the assumed $f(n)$.

These facts suggest that the HCDs in the data are either more numerous or more biased than in the mocks
or that there are additional effects in the data that smear the correlations in the radial direction, which are then modeled with the HCD parameters. 
The former seems unlikely since the DLA content of the data is well-known \citep{noterdaeme2012column} and even in the masking case this study indicates that most of the HCD mode-suppression comes from HCDs with values of $n$ just below that of DLAs.
The possibility that there are other mode-suppressing effects then justifies the continued use of the \Expmodel~model in the DESI analyses.

\section{Conclusion and Discussion}\label{sec:conclusions}
We have applied the \Voimodel~model proposed in MW12 and FM12
to characterize the effect of HCDs on the Ly$\alpha$ forest correlation function and power spectrum. This model is based on the Voigt absorption profile that parametrizes the damping wings of HCDs, and takes into account the HI column density probability distribution of HCDs, and the biases of the halos containing the HCD's. Note that in this work, we do not account for the redshift dependence. This approximation is appropriate for the current analysis, where the Ly$\alpha$ forest correlation functions are measured at an effective redshift. However, a more detailed treatment of the redshift dependence will be necessary in future analyses.

We performed a suite of verifications of the model, based on the fitting to Ly$\alpha$ forest correlations, computed from mocks with different HI column density probability distributions of HCDs. 
The fitted values of $\bhcd$ are generally consistent with the expected value $\bhcd=2$, confirming the validity of the model.
Compared with the adhoc $\texttt{Exp}$ model,
the $\texttt{Voigt}$ model gives comparable $\chi^2$, with a smaller uncertainty for $b_{\eta\text{LY}\alpha}$ and $\beta_{\text{Ly}\alpha}$. 

Application of the \Voimodel~model to real data will be challenging
because of uncertainties in the HCD content and bias and the effects of masking false DLA detections.
If it can be understood,
the HI column density distribution in the range of $17<n<20$, could also potentially be constrained with this model, which is difficult to measure from direct observation.
However, the large value of $\bhcd$ fit with real data suggests that other smearing effects are present in the data.
Fortunately, the eBOSS DR16 data confirms that the modeling of HCDs does not affect the measurement of BAO significantly. Validation analyses based on DESI data will be further investigated for this model.

\section{Data availability}
The data points corresponding to each figure in this paper can be accessed in the Zenodo
repository at \href{https://doi.org/10.5281/zenodo.15591451}{https://doi.org/10.5281/zenodo.15591451}.

\appendix

\section{Derivation of the \Voimodel~and \Expmodel~models}

Following the work of MW11 and FM12,
we derive here eqn. \ref{eqn:scale-dependent_bias}, found in this appendix as eqns. \ref{eq:biasTkpar} and \ref{eq:biasetaTkpar}.
The motivation is to  correct some minor errors in these papers and in \citep{rogers2018correlations}.


We divide the absorption caused by neutral hydrogen in two groups: the absorption $\tau_\alpha$ caused by the \lyaf in the Intergalactic Medium (IGM) and the absorption $\tau_H$ caused by High Column Density systems (HCDs) in the vicinity of galaxies.
The total transmitted flux fraction $F_T$  due to absorption at the position $x$ will be the product of both:
\begin{align}
F_T(x) & = F_\alpha(x) F_H(x)  \\
       & = \overline{F_\alpha} \left[ 1 + \delta_\alpha(x) \right] 
           \overline{F_H} \left[ 1 + \delta_H(x) \right]      \\ \nonumber
       & = \overline{F_T} \left[ 1 + \delta_T(x) \right]  ~.     \nonumber
\end{align}
Since $\delta_\alpha$ and $\delta_H$ are correlated, the relation between the mean fluxes is:
\begin{equation}
 \overline{F_T}  = \left< F_T \right> = \overline{F_\alpha} \; \overline{F_H} \left[1 + C \right]~,
\end{equation}
where we have introduced $C = \left< \delta_\alpha(x) \delta_H(x) \right>$.

The fluctuations in the total transmitted flux fraction are then:
\begin{equation}
 1 + \delta_T(x) = \frac{1}{1+C} \left[1+\delta_\alpha(x)\right] 
    \left[1+\delta_H(x)\right] ~,
\end{equation}
or in other words,
\begin{equation}
\delta_T = \frac{\delta_\alpha + \delta_H + \delta_\alpha \delta_H - C }{1+C} ~.
\end{equation}

FM12 estimate $C\approx0.003$ and
for the rest of the discussion we will make the approximation $C=0$ and  ignore the second order term $\delta_\alpha \delta_H$, obtaining the simple relation 
\begin{equation}
\delta_T \approx \delta_\alpha + \delta_H  ~.
\end{equation}
{If we had not made this approximation
we should not obtain that the total flux bias is the sum of that due to the IGM and that due to the HCDs, since this ignores $C$ and the other higher order terms that would modify even the linear bias parameters.}

\subsection{Absorption due to the Voigt profile}

The absorption at radial position $x$ due to the presence of a HCD  of column density $\nhi$ at $x^\prime$ is
\begin{equation}
1 - F_H(x) = \left[1 - V(x^\prime-x, n)\right] ~,
\label{eqn:defineV}\end{equation}
where $V(x^\prime -x, n=\lognhi)$ is a Voigt profile for a column density $\nhi$, normalized at $V=1$ at large $x^\prime -x$.\footnote{This definition of $V$ corresponds to $(1-V)$ of FM12}
The probability of having a HCD  at a given position $x^\prime$ and column density $\nhi$ is 
$\rhohcd f(n)dx^\prime dn$
where $\rhohcd$ is the density of HCDs and $f(n)$ is the normalized distribution of $n=\lognhi$.
In a region with a linear overdensity $\delta_L(x^\prime)$, the probability is enhanced
\begin{equation}
 p(n, x^\prime ) = \rhohcd f(n) \left[ 1 + \bhcd (n) ~ \delta_L(x^\prime ) \right] ~,
\end{equation}
where we have introduced the linear bias of the halos hosting HCDs, in principle a function of column density $\bhcd (n)$.

Using this probability we can write the integrals:
\begin{equation}
1 - F_H(x) = \int dx^\prime \int dn \; p(n, x^\prime) \left[1 - V(x^\prime-x, n) \right] ~,
\end{equation}
and compute the value of $\overline{F_H}$ using the value of $p(n,x^\prime )$ at $\delta_L=0$:
\begin{equation}
1 - \overline{F_H} 
 = \rhohcd\int dn f(n) ~ W(n) ~,   
\label{eq:oneminusFhcdmean}
\end{equation}
where we have defined 
\begin{equation}
 W(n) = \int dx^\prime \left[1 - V(x^\prime-x, n) \right] ~,
\end{equation}
the equivalent width of the Voigt profile in comoving units (not in \AA).

We can now describe the fluctuation in HCD absorption as:
\begin{align}
 \delta_H(x) &= \frac{F_H(x) - \overline{F_H}}{\overline{F_H}}     \\
  &= -\frac{1}{\overline{F_H}} \int dx^\prime \int dn \left[ p(n, x^\prime) - \rhohcd f(n) \right] \left[1 - V(x^\prime-x, n) \right]      \\ \nonumber
  &= -\frac{\rhohcd}{\overline{F_H}} \int dn f(n) \bhcd (n) \int dx^\prime \delta_L(x^\prime) \left[1 - V(x^\prime-x, n) \right]      \nonumber    ~.
\end{align}

The expression is clearer in Fourier space, since the convolution becomes a product of Fourier modes:
\begin{equation}
 \delta_H(\kpar ) = -\rhohcd\frac{\delta_L(\kpar )}{\overline{F_H}} \int dn f(n) ~ \bhcd (n) ~ W(\kpar , n) ~,
 \label{eqn:deltaHofk}
\end{equation}
where $W(\kpar ,n)$ is the Fourier transform of $\left[ 1-V(x,n) \right]$.

\subsection{Linear bias of HCD absorption}

On very large scales, larger than the typical size of DLAs, we should recover a linear bias relation. 
To distinguish between the linear bias of the HCD systems ($b_{HCD}$) and the linear bias of the absorption caused by these, we will refer to the latter as $b_{HCD}^F$, i.e., $\delta_H(k) = b_{HCD}^F \delta_L(k)$. 
Using eqn. \ref{eqn:deltaHofk} we get:
\begin{equation}
 \bfhcd =
  \frac{\rhohcd}{\overline{F_H}} \int dn f(n) ~ \bhcd (n) ~ W(n) ~,     
  \label{eq:bfhcdvoigt}
\end{equation}
where we have used the fact that 
\begin{align}
 \lim_{\kpar \to 0} W(\kpar , n) 
  &= \lim_{\kpar \to 0} \int dx ~ e^{i\kpar x} \left[1 - V(x, n\right)]      \\
  &= \int dx \left[1 - V(x, n\right)] = W(n) ~.        \nonumber 
\end{align}


For a $n$-independent bias $\bhcd(N)=\bhcd$, using eqn. \ref{eq:oneminusFhcdmean} this gives
\begin{equation}
 \bfhcd = \bhcd \frac{1 - \overline{F_H}}{\overline{F_H}} ~.
 \label{eq:bfhcNindependent}
\end{equation}

\subsection{Relation with the Exponential model}

In the \Expmodel~model used in previous eBOSS \citep{dmdb2020} and DESI \citep{DESIlya} analyses, 
we describe the contamination as
\begin{equation}
 \delta_H(k) = \bfhcd ~ \exp[-\Lhcd \kpar]
 ~ \delta_L(k) ~,
\end{equation}
where $\bfhcd$ is the large-scale bias.
Using eqns. 
\ref{eqn:deltaHofk} and \ref{eq:bfhcdvoigt}, we can write a similar equation for the \Voimodel~model:
\begin{equation}
 \delta_H(k) = \bfhcd ~ F_V(\kpar) ~ \delta_L(k) ~,
\end{equation}
where $\bfhcd$ is the linear bias discussed above, and we have introduced
\begin{equation}
 \label{eq:F_V}
 F_V(\kpar) = \frac{\int dn f(n) ~ \bhcd(n) ~ W(\kpar, n)}{\int dn f(n) ~ \bhcd(n) ~ W(n)}~.
\end{equation}
The correspondence with $\Fhcdvoigt$  in eqn. \ref{Equa:Fhcd} is
\begin{equation}
    \bhcd \Fhcdvoigt(\kpar) = \bfhcd F_V(\kpar)
\end{equation}

\subsection{Redshift space distortions}

To include redshift-space distortions, we replace $\bhcd$ with $\bhcd + f\mu_k^2$ where
$f\approx1$ (no argument $n$) is the linear growth rate and $\mu_k=\kpar/k$ with $\kpar$ being
the component of $k$ along the line of sight.
%
\begin{equation}
 \delta_H(k, \mu_k) = \rhohcd \frac{\delta_L(k)}{\overline{F_H}} \int dn f(n) \left[\bhcd(n) + f \mu_k^2 \right] W(\kpar, n) ~.
\end{equation}

Similarly to how we have computed $\bfhcd$ above, we can now compute the low-k limit for $\mu_k=1$:
\begin{align}
 \bfhcd \left[ 1 + \betahcd \right]  
  &= \lim_{k \to 0} \frac{\delta_H(k, \mu_k=1)}{\delta_L(k)}      \\
  &= \frac{\rhohcd}{\overline{F_H}} \int dn f(n) ~ \left[\bhcd(n) + f \right] ~ W(n) \\
  &= \bfhcd + f \frac{1 - \overline{F_H}}{\overline{F_H}} ~.
\end{align}

From this, one can obtain the value of $\betahcd$:
\begin{equation}
 \betahcd = f \frac{1}{\bfhcd} \frac{1-\overline{F_H}}{\overline{F_H}} ~.
\end{equation}

Whenever we have $\bhcd(n)=\bhcd$ we can 
further simplify the expression:
\begin{equation}
 \betahcd = \frac{f}{\bhcd} ~.
\end{equation}

\subsection{Contaminated Kaiser model}

Assuming again that $\delta_T = \delta_\alpha + \delta_H$ and $\bfhcd(n)=\bfhcd$, it is easy to see that 
\begin{equation}
 \frac{\delta_T(k,\mu_k)}{\delta_L(k)} 
   = \left[ \blya + \bfhcd F_V(\kpar) \right] 
   + \left[ \blya \betalya + \bfhcd \betahcd F_V(\kpar) \right] \mu_k^2 ~,
\end{equation}
i.e.,
\begin{equation}
 b_T(\kpar) = \blya + \bfhcd F_V(\kpar) 
\label{eq:biasTkpar}
\end{equation}
and
\begin{equation}
 b^\eta_T (\kpar) = \biasetalya + \biasetaHCD F_V(\kpar) ~,
 \label{eq:biasetaTkpar}
\end{equation}
where we have used the \textit{velocity gradient bias}
\begin{equation}
b^\eta_X = \frac{b_X \beta_X}{f} ~.
\end{equation}
Equations \ref{eq:biasTkpar}
and \ref{eq:biasetaTkpar}
are equivalent to equation \ref{eqn:scale-dependent_bias} with the identification $\bfhcd F_V(\kpar)=\bhcd \Fhcdvoigt(\kpar)$.

\subsection{Comparison to MW11}

Eq. (B.1) of MW11 shows the power spectrum of HCD contamination:
\begin{equation}
 P_H(k,\mu) = \left(b_h + \mu^2 \right)^2 P_L(k) \left(\tilde W_2(\kpar) \right)^2 ~,
\end{equation}
where the authors assume that all HCDs have the same bias and that $f=1$.
Their $b_h$ corresponds to our $\bhcd$.

The authors define $\tilde W_2(\kpar)$ in Eq. (B.2) as:
\begin{equation}
 \tilde W_2(\kpar) = \int dN f(N) W(N,k) ~,
\end{equation}
(although they use $\tilde{d}(\kpar)$ to refer to $W(k,N)$).
Their $f(N)$ corresponds to our $\rhohcd f(n)$

On the other hand, our model predicts that the power of HCD contamination should be:
\begin{equation}
 P_H(k,\mu) = \left(\frac{b_H}{b_h}\right)^2 \left(b_h + f \mu^2 \right)^2 P_L(k) ~ F_V^2(\kpar) ~.
\end{equation}

If we assume again that $b_h(N)=b_h$, we can simplify the equation using
\begin{align}
 \frac{b_H}{b_h} F_V(\kpar) 
   & = \frac{1-\overline{F_H}}{\overline{F_H}} F_V(k)        \\
   & = \frac{1-\overline{F_H}}{\overline{F_H}} 
       \frac{\int dN f(N) ~ W(k, N)}{\int dN f(N) ~ W(N)}    \\ \nonumber
   & = \frac{1}{\overline{F_H}} \int dN f(N) ~ W(k, N)       \\ \nonumber
   & = \frac{1}{\overline{F_H}} \tilde{W_2}(k)      ~.          \nonumber
\end{align}

It appears that MW11 missed a factor of $1/\overline{F_H}$ in their calculations. This number is very close to one, so the difference is minor.

\subsection{Comparison to FM12}

Eq. (4.25) in FM12 shows the relation between the cross-correlation term $\xi^V_{\alpha H}$ taking into account the Voigt profile, compared to the same term without the Voigt profile $\xi_{\alpha H}$.
The authors used $V(N)$ to refer to the absorption, i.e., what here we call $1-V(N)$. 

If we do the Fourier transform, and rewrite it in terms of comoving coordinates (and not velocities), we can rewrite the equation as:
\begin{equation}
 P^V_{\alpha H}(k, \mu) = F_V(\kpar) P_{\alpha H}(k, \mu) ~,
\end{equation}
with 
\begin{equation}
 F_V(\kpar) = \frac{\int dN f(N) ~ W(N,k)}{ \int dN f(N) ~ W(N)} ~.
\end{equation}

This is similar to our Eq. \ref{eq:F_V}, but only if one assumes $b_h(N)=b_h$.
This is not important when analyzing mocks, but in general the halo bias could be a function of column density.

\subsection{Comparison to Rogers et al. (2018)}

Eq. (3) in \citet{rogers2018correlations} shows the estimate for $F_V(\kpar)$ to be:
\begin{equation}
 F_V(k) = \int dN f(N) ~ W(N, k) ~.
\end{equation}

This is significantly different than our results, since they ignore the normalization factor, so their $F_V(k)$ does not go to 1 on large scales, and it goes instead to $1-\overline{F_H}$.

The results would be equivalent to \cite{McQuinn2011} if $b_{\rm HCD}$ in Eq. (2) of \cite{rogers2018correlations} meant the halo bias (our $\bhcd$), not the absorption bias (our $\bfhcd$) as stated.

\section{Complete results of fits}

Tables \ref{Tab:voigt_fits_unique_n}, \ref{Tab:fits_realistic_n}, \ref{Tab:fitsbinsize2}, and \ref{Tab:fits_eboss} give the complete results of fits on the mocks and dr16 eBOSS data.

\begin{table*}
\begin{center}
\scalebox{0.85}{\begin{tabular}{|c|c|c|c|c|}

\hline
&$\rm{Ly}\alpha\times\rm{Ly}\alpha$ & $\rm{Ly}\alpha\times\rm{Ly}\alpha$& $\rm{Ly}\alpha\times\rm{Ly}\alpha$ & $\rm{Ly}\alpha\times\rm{Ly}\alpha$ \\
$n$&19.5&20.0&20.5&21.0\\ 
$N_{\rm{mocks}}$&10&10&10&10 \\ 
$z_{\rm{eff}}$&2.29&2.29&2.29&2.29 \\ 
$[r_{\rm{min}},r_{\rm{max}}](h^{-1}\rm{Mpc})$&[20,180]&[20,180]&[20,180]&[20,180] \\ 
$\chi^2$&1583.58&1584.23&1583.63&1682.56\\ 
$N_{\rm{data}}$&1574&1574&1574&1574 \\ 
$N_{\rm{par}}$&6&6&6&6\\ 
$P$&0.39&0.38&0.39&0.02\\ 
$\alpha_{||}$&1.01$\pm$0.009&1.01$\pm$0.009&1.01$\pm$0.01&1.0$\pm$0.01 \\ 
$\alpha_{\perp}$&0.998$\pm$0.013&1.0$\pm$0.015&1.0$\pm$0.017&1.01$\pm$0.019 \\ 
$b_{\eta,\rm{Ly}\alpha}$&-0.206$\pm$0.002&-0.204$\pm$0.003&-0.201$\pm$0.003&-0.187$\pm$0.003 \\ 
$\beta_{\rm{Ly}\alpha}$&1.61$\pm$0.04&1.6$\pm$0.05&1.57$\pm$0.06&1.33$\pm$0.05 \\ 
$b_{\rm{HCD}}$&1.95$\pm$0.292&2.13$\pm$0.157&2.2$\pm$0.092&1.92$\pm$0.052 \\ 
$\beta_{\rm{HCD}}$&0.48$\pm$0.09&0.48$\pm$0.08&0.5$\pm$0.07&0.75$\pm$0.05 \\ 
\hline
\\
\hline
&$\rm{Ly}\alpha\times\rm{QSO}$ & $\rm{Ly}\alpha\times\rm{QSO}$& $\rm{Ly}\alpha\times\rm{QSO}$ & $\rm{Ly}\alpha\times\rm{QSO}$\\
$z_{\rm{eff}}$&2.29&2.29&2.29&2.29 \\ 
$[r_{\rm{min}},r_{\rm{max}}](h^{-1}\rm{Mpc})$&[40,180]&[40,180]&[40,180]&[40,180] \\ 
$\chi^2$&2994.79&3029.07&3056.4&3074.56 \\ 
$N_{\rm{data}}$&3030&3030&3030&3030 \\ 
$N_{\rm{par}}$&7&7&7&7 \\ 
$P$&0.64&0.47&0.33&0.25 \\ 
$\alpha_{||}$&1.0$\pm$0.009&1.0$\pm$0.01&1.0$\pm$0.01&0.998$\pm$0.01 \\ 
$\alpha_{\perp}$&1.0$\pm$0.011&1.0$\pm$0.012&1.0$\pm$0.013&1.0$\pm$0.014 \\ 
$b_{\eta,\rm{Ly}\alpha}$&-0.194$\pm$0.004&-0.191$\pm$0.006&-0.19$\pm$0.007&-0.191$\pm$0.007 \\ 
$\beta_{\rm{Ly}\alpha}$&1.67$\pm$0.08&1.78$\pm$0.13&1.89$\pm$0.18&1.67$\pm$0.14 \\ 
$b_{\rm{HCD}}$&2.3$\pm$0.488&2.64$\pm$0.427&2.59$\pm$0.275&2.12$\pm$0.138 \\ 
$\beta_{\rm{HCD}}$&0.51$\pm$0.09&0.53$\pm$0.09&0.56$\pm$0.08&0.63$\pm$0.08 \\ 
$\Delta r_{||,\rm{QSO}}$ $(h^{-1}\rm{Mpc})$&0.41$\pm$0.17&0.36$\pm$0.18&0.27$\pm$0.2&0.19$\pm$0.22 \\ 
\hline

\end{tabular}}
\caption{
Best fit parameters for stack of 10 mocks created with HCDs with the column densities $n=\lognhi=19.5, 20.0, 20.5, 21.0$, using the \texttt{Voigt} model, for Ly$\alpha$ auto-correlation function and Ly$\alpha$-quasar cross-correlation.
}
\label{Tab:voigt_fits_unique_n}
\end{center}
\end{table*}

\begin{table*}
\begin{center}
\scalebox{0.85}{\begin{tabular}{|c|c|c|c|c|}

\hline
$\rm{Model}$&$\rm{Exp}$&\rm{Voigt}&\rm{Exp}&\rm{Voigt} \\ 
$\rm{Masking}$&$\rm{Yes}$&\rm{Yes}&\rm{No}&\rm{No} \\ 
&$\rm{Ly}\alpha\times\rm{Ly}\alpha$ & $\rm{Ly}\alpha\times\rm{Ly}\alpha$& $\rm{Ly}\alpha\times\rm{Ly}\alpha$ & $\rm{Ly}\alpha\times\rm{Ly}\alpha$ \\
$N_{\rm{mocks}}$&10&10&10&10 \\ 
$z_{\rm{eff}}$&2.29&2.29&2.29&2.29 \\ 
$[r_{\rm{min}},r_{\rm{max}}](h^{-1}\rm{Mpc})$&[20,180]&[20,180]&[20,180]&[20,180] \\ 
$\chi^2$&1522.52&1523.24&1533.47&1534.13\\ 
$N_{\rm{data}}$&1574&1574&1574&1574 \\ 
$N_{\rm{par}}$&7&6&7&6\\ 
$P$&0.79&0.79&0.72&0.72\\ 
$\alpha_{||}$&1.01$\pm$0.009&1.01$\pm$0.009&1.0$\pm$0.009&1.0$\pm$0.009 \\ 
$\alpha_{\perp}$&0.98$\pm$0.013&0.98$\pm$0.013&0.984$\pm$0.015&0.984$\pm$0.015 \\ 
$b_{\eta,\rm{Ly}\alpha}$&-0.209$\pm$0.002&-0.206$\pm$0.001&-0.208$\pm$0.002&-0.205$\pm$0.002 \\ 
$\beta_{\rm{Ly}\alpha}$&1.56$\pm$0.04&1.64$\pm$0.04&1.58$\pm$0.05&1.64$\pm$0.04 \\ 
$\beta_{\rm{HCD}}$&0.49$\pm$0.09&0.48$\pm$0.09&0.48$\pm$0.09&0.48$\pm$0.08 \\ 
$b_{\rm{HCD}}$&&1.93$\pm$0.346&&1.82$\pm$0.146 \\ 
$b^F_{\rm{HCD}}$&-0.008$\pm$0.004&&-0.026$\pm$0.004& \\ 
$L_{\rm{HCD}}$&7.86$\pm$6.64&&9.5$\pm$2.56& \\ 
\hline
\\
\hline
&$\rm{Ly}\alpha\times\rm{QSO}$ & $\rm{Ly}\alpha\times\rm{QSO}$& $\rm{Ly}\alpha\times\rm{QSO}$ & $\rm{Ly}\alpha\times\rm{QSO}$\\
$z_{\rm{eff}}$&2.29&2.29&2.29&2.29 \\ 
$[r_{\rm{min}},r_{\rm{max}}](h^{-1}\rm{Mpc})$&[40,180]&[40,180]&[40,180]&[40,180] \\ 
$\chi^2$&3133.7&3138.72&3110.47&3112.57 \\ 
$N_{\rm{data}}$&3030&3030&3030&3030 \\ 
$N_{\rm{par}}$&8&7&8&7 \\ 
$P$&0.08&0.07&0.13&0.13 \\ 
$\alpha_{||}$&1.0$\pm$0.009&1.0$\pm$0.009&1.0$\pm$0.01&1.0$\pm$0.009 \\ 
$\alpha_{\perp}$&1.0$\pm$0.011&1.0$\pm$0.011&0.996$\pm$0.012&0.997$\pm$0.012 \\ 
$b_{\eta,\rm{Ly}\alpha}$&-0.169$\pm$0.015&-0.193$\pm$0.003&-0.176$\pm$0.034&-0.191$\pm$0.004 \\ 
$\beta_{\rm{Ly}\alpha}$&2.83$\pm$0.92&1.63$\pm$0.05&2.52$\pm$1.68&1.78$\pm$0.11 \\ 
$\beta_{\rm{HCD}}$&0.51$\pm$0.07&0.51$\pm$0.09&0.52$\pm$0.09&0.53$\pm$0.09 \\ 
$b_{\rm{HCD}}$&&2.2$\pm$0.49&&2.46$\pm$0.387 \\ 
$b^F_{\rm{HCD}}$&-0.077$\pm$0.024&&-0.082$\pm$0.057& \\ 
$L_{\rm{HCD}}$&2.56$\pm$1.22&&4.96$\pm$4.44& \\ 
$\Delta r_{||,\rm{QSO}}$ $(h^{-1}\rm{Mpc})$&0.2$\pm$0.17&0.2$\pm$0.17&0.18$\pm$0.19&0.19$\pm$0.19 \\ 
\hline

\end{tabular}}
\caption{
Best fit parameters for stack of 10 mocks created with HCDs with \texttt{pyigm} column densities, using the \texttt{Voigt}~and \Expmodel~models, for Ly$\alpha$ auto-correlation function and Ly$\alpha$-quasar cross-correlation, respectively.
}
\label{Tab:fits_realistic_n}
\end{center}
\end{table*}

\begin{table*}
\begin{center}
\scalebox{0.85}{\begin{tabular}{|c|c|c|c|c|}

\hline
$\rm{Model}$&$\rm{Exp}$&\rm{Voigt}&\rm{Exp}&\rm{Voigt} \\ 
$\rm{Masking}$&$\rm{Yes}$&\rm{Yes}&\rm{No}&\rm{No} \\ 
&$\rm{Ly}\alpha\times\rm{Ly}\alpha$ & $\rm{Ly}\alpha\times\rm{Ly}\alpha$& $\rm{Ly}\alpha\times\rm{Ly}\alpha$ & $\rm{Ly}\alpha\times\rm{Ly}\alpha$ \\
$N_{\rm{mocks}}$&10&10&10&10 \\ 
$z_{\rm{eff}}$&2.29&2.29&2.29&2.29 \\ 
$[r_{\rm{min}},r_{\rm{max}}](h^{-1}\rm{Mpc})$&[20,180]&[20,180]&[20,180]&[20,180] \\ 
$\chi^2$&6340.93&6344.04&6308.51&6309.75\\ 
$N_{\rm{data}}$&6048&6048&6048&6048 \\ 
$N_{\rm{par}}$&7&6&7&6\\ 
$P$&0.0&0.0&0.01&0.01\\ 
$\alpha_{||}$&1.0$\pm$0.009&1.0$\pm$0.009&1.0$\pm$0.009&1.0$\pm$0.009 \\ 
$\alpha_{\perp}$&0.984$\pm$0.014&0.983$\pm$0.014&0.984$\pm$0.015&0.984$\pm$0.015 \\ 
$b_{\eta,\rm{Ly}\alpha}$&-0.206$\pm$0.001&-0.202$\pm$0.001&-0.205$\pm$0.002&-0.202$\pm$0.002 \\ 
$\beta_{\rm{Ly}\alpha}$&1.48$\pm$0.02&1.56$\pm$0.03&1.53$\pm$0.05&1.58$\pm$0.04 \\ 
$\beta_{\rm{HCD}}$&0.48$\pm$0.09&0.46$\pm$0.09&0.47$\pm$0.09&0.47$\pm$0.09 \\ 
$b_{\rm{HCD}}$&&1.58$\pm$0.355&&1.66$\pm$0.149 \\ 
$b^F_{\rm{HCD}}$&-0.004$\pm$0.002&&-0.024$\pm$0.004& \\ 
$L_{\rm{HCD}}$&15.09$\pm$11.39&&9.12$\pm$2.79& \\ 
\hline
\\
\hline
&$\rm{Ly}\alpha\times\rm{QSO}$ & $\rm{Ly}\alpha\times\rm{QSO}$& $\rm{Ly}\alpha\times\rm{QSO}$ & $\rm{Ly}\alpha\times\rm{QSO}$\\
$z_{\rm{eff}}$&2.29&2.29&2.29&2.29 \\ 
$[r_{\rm{min}},r_{\rm{max}}](h^{-1}\rm{Mpc})$&[40,180]&[40,180]&[40,180]&[40,180] \\ 
$\chi^2$&12209.06&12215.94&12290.08&12292.53 \\ 
$N_{\rm{data}}$&12092&12092&12092&12092 \\ 
$N_{\rm{par}}$&8&7&8&7 \\ 
$P$&0.21&0.2&0.09&0.09 \\ 
$\alpha_{||}$&1.0$\pm$0.009&1.0$\pm$0.01&1.01$\pm$0.01&1.0$\pm$0.01 \\ 
$\alpha_{\perp}$&0.999$\pm$0.012&0.999$\pm$0.012&0.996$\pm$0.013&0.997$\pm$0.013 \\ 
$b_{\eta,\rm{Ly}\alpha}$&-0.2$\pm$0.003&-0.191$\pm$0.003&-0.18$\pm$0.009&-0.189$\pm$0.004 \\ 
$\beta_{\rm{Ly}\alpha}$&1.6$\pm$0.08&1.65$\pm$0.06&2.3$\pm$0.31&1.79$\pm$0.11 \\ 
$\beta_{\rm{HCD}}$&0.51$\pm$0.09&0.51$\pm$0.09&0.52$\pm$0.08&0.53$\pm$0.09 \\ 
$b_{\rm{HCD}}$&&2.21$\pm$0.49&&2.5$\pm$0.385 \\ 
$b^F_{\rm{HCD}}$&-0.016$\pm$0.005&&-0.074$\pm$0.012& \\ 
$L_{\rm{HCD}}$&32.34$\pm$33.3&&5.87$\pm$1.69& \\ 
$\Delta r_{||,\rm{QSO}}$ $(h^{-1}\rm{Mpc})$&0.49$\pm$0.18&0.49$\pm$0.18&0.52$\pm$0.19&0.53$\pm$0.19 \\ 
\hline

\end{tabular}}
\caption{Same as Table \ref{Tab:fits_realistic_n} except now with binning size of $2h^{-1}\text{Mpc}$.
}
\label{Tab:fitsbinsize2}
\end{center}
\end{table*}

\begin{table*}
\begin{center}
\scalebox{0.85}{\begin{tabular}{|c|c|c|c|c|}

\hline
$\rm{Model}$&$\rm{Exp}$&\rm{Voigt}&\rm{Exp}&\rm{Voigt} \\ 
$\rm{Masking}$&$\rm{Yes}$&\rm{Yes}&\rm{No}&\rm{No} \\ 
&$\rm{Ly}\alpha\times\rm{Ly}\alpha$ & $\rm{Ly}\alpha\times\rm{Ly}\alpha$& $\rm{Ly}\alpha\times\rm{Ly}\alpha$ & $\rm{Ly}\alpha\times\rm{Ly}\alpha$ \\
$z_{\rm{eff}}$&2.29&2.29&2.29&2.29 \\ 
$[r_{\rm{min}},r_{\rm{max}}](h^{-1}\rm{Mpc})$&[10,180]&[10,180]&[10,180]&[10,180] \\ 
$\chi^2$&1576.22&1624.34&1594.96&1630.65\\ 
$N_{\rm{data}}$&1590&1590&1590&1590 \\ 
$N_{\rm{par}}$&14&13&14&13\\ 
$P$&0.49&0.2&0.36&0.17\\ 
$\alpha_{||}$&1.05$\pm$0.034&1.04$\pm$0.033&1.04$\pm$0.034&1.04$\pm$0.033 \\ 
$\alpha_{\perp}$&0.981$\pm$0.042&0.985$\pm$0.041&0.974$\pm$0.044&0.973$\pm$0.044 \\ 
$b_{\eta,\rm{Ly}\alpha}$&-0.175$\pm$0.013&-0.179$\pm$0.004&-0.173$\pm$0.013&-0.189$\pm$0.005 \\ 
$\beta_{\rm{Ly}\alpha}$&3.23$\pm$1.26&1.71$\pm$0.11&5.25$\pm$3.29&1.84$\pm$0.14 \\ 
$\beta_{\rm{HCD}}$&0.53$\pm$0.08&0.67$\pm$0.08&0.51$\pm$0.08&0.67$\pm$0.08 \\ 
$b_{\rm{HCD}}$&&7.3$\pm$0.611&&4.79$\pm$0.326 \\ 
$b^F_{\rm{HCD}}$&-0.105$\pm$0.022&&-0.139$\pm$0.02& \\ 
$L_{\rm{HCD}}$&2.28$\pm$0.63&&2.59$\pm$0.52& \\ 
\hline
\\
\hline
&$\rm{Ly}\alpha\times\rm{QSO}$ & $\rm{Ly}\alpha\times\rm{QSO}$& $\rm{Ly}\alpha\times\rm{QSO}$ & $\rm{Ly}\alpha\times\rm{QSO}$\\
$z_{\rm{eff}}$&2.29&2.29&2.29&2.29 \\ 
$[r_{\rm{min}},r_{\rm{max}}](h^{-1}\rm{Mpc})$&[10,180]&[10,180]&[10,180]&[10,180] \\ 
$\chi^2$&3220.27&3221.4&3219.44&3224.94 \\ 
$N_{\rm{data}}$&3180&3180&3180&3180 \\ 
$N_{\rm{par}}$&13&12&13&12 \\ 
$P$&0.25&0.25&0.25&0.24 \\ 
$\alpha_{||}$&1.06$\pm$0.032&1.06$\pm$0.032&1.05$\pm$0.034&1.05$\pm$0.034 \\ 
$\alpha_{\perp}$&0.932$\pm$0.039&0.933$\pm$0.039&0.948$\pm$0.042&0.947$\pm$0.042 \\ 
$b_{\eta,\rm{Ly}\alpha}$&-0.228$\pm$0.016&-0.237$\pm$0.014&-0.231$\pm$0.019&-0.27$\pm$0.018 \\ 
$\beta_{\rm{Ly}\alpha}$&1.91$\pm$0.33&1.91$\pm$0.21&1.92$\pm$0.34&1.56$\pm$0.16 \\ 
$\beta_{\rm{HCD}}$&0.52$\pm$0.09&0.51$\pm$0.09&0.51$\pm$0.09&0.5$\pm$0.09 \\ 
$b_{\rm{HCD}}$&&3.78$\pm$1.92&&-0.424$\pm$1.4 \\ 
$b^F_{\rm{HCD}}$&-0.034$\pm$0.024&&-0.047$\pm$0.027& \\ 
$L_{\rm{HCD}}$&0.95$\pm$2.87&&-0.01$\pm$1.66& \\ 
$\Delta r_{||,\rm{QSO}}$ $(h^{-1}\rm{Mpc})$&0.15$\pm$0.13&0.19$\pm$0.13&0.19$\pm$0.13&0.21$\pm$0.13 \\ 
$\sigmav$ $(h^{-1}\rm{Mpc})$&8.94$\pm$0.69&8.94$\pm$0.55&9.79$\pm$0.78&10.3$\pm$0.84 \\ 
\hline

\end{tabular}}
\caption{Best fit parameters for eBOSS DR16 data, using the \texttt{Voigt} model and the \texttt{Exp} model, for Ly$\alpha$ auto-correlation function and Ly$\alpha$-quasar cross-correlation, respectively.}
\label{Tab:fits_eboss}
\end{center}
\end{table*}

\acknowledgments
This material is based upon work supported by the U.S. Department of Energy (DOE), Office of Science, Office of High-Energy Physics, under Contract No. DE–AC02–05CH11231, and by the National Energy Research Scientific Computing Center, a DOE Office of Science User Facility under the same contract. Additional support for DESI was provided by the U.S. National Science Foundation (NSF), Division of Astronomical Sciences under Contract No. AST-0950945 to the NSF’s National Optical-Infrared Astronomy Research Laboratory; the Science and Technology Facilities Council of the United Kingdom; the Gordon and Betty Moore Foundation; the Heising-Simons Foundation; the French Alternative Energies and Atomic Energy Commission (CEA); the National Council of Humanities, Science and Technology of Mexico (CONAHCYT); the Ministry of Science, Innovation and Universities of Spain (MICIU/AEI/10.13039/501100011033), and by the DESI Member Institutions: \url{https://www.desi.lbl.gov/collaborating-institutions}. Any opinions, findings, and conclusions or recommendations expressed in this material are those of the author(s) and do not necessarily reflect the views of the U. S. National Science Foundation, the U. S. Department of Energy, or any of the listed funding agencies.

The authors are honored to be permitted to conduct scientific research on I'oligam Du'ag (Kitt Peak), a mountain with particular significance to the Tohono O’odham Nation.

Ting Tan thanks the French CNRS International PhD program for its support. The authors thank Andreu Font-Ribera for his contribution of writing Appendix A of this paper. We also thank Julien Guy for his contribution to the derivation of the model included in this paper.


\bibliographystyle{unsrtnat_arxiv}  
\bibliography{reference}		

\end{document}